\documentclass{siamltex}

\usepackage{amsfonts,amsmath,graphicx,subfigure}
\usepackage{verbatim}
\usepackage{siunitx}
\usepackage{color}
\usepackage{adjustbox}

\newcommand{\JMHPBB}[1]{\textcolor{black}{{#1}}}
\newcommand{\JMHREV}[1]{\textcolor{black}{{#1}}}
\newcommand{\JMH}[1]{\textcolor{black}{{#1}}}

\newtheorem{remark}{Remark}

\addcontentsline{toc}{chapter}{
Learning Compact Physics-Aware Delayed Photocurrent Models Using Dynamic Mode Decomposition\\
{\em J.\ Hanson, P.\ Bochev, and B.\ Paskaleva}}

\title{Learning Compact Physics-Aware \JMHPBB{Delayed} Photocurrent Models Using Dynamic Mode Decomposition\protect\thanks{Sandia National Laboratories is a multimission laboratory managed and operated by National Technology and Engineering Solutions of Sandia, LLC., a wholly owned subsidiary of Honeywell International, Inc., for the U.S. Department of Energy’s National Nuclear Security Administration under contract DE-NA-0003525.}}

\author{
Joshua Hanson\thanks{University of Illinois at Urbana-Champaign, jmh4@illinois.edu} \and
Pavel Bochev\thanks{Sandia National Laboratories, pbboche@sandia.gov} \and
Biliana Paskaleva\thanks{Sandia National Laboratories, bspaska@sandia.gov}}

\begin{document}

\maketitle

\begin{abstract}
Radiation-induced photocurrent in semiconductor devices \JMHPBB{can be} simulated using complex physics-based models, which are accurate, but computationally expensive. This presents a challenge for implementing device characteristics in high-level circuit simulations where it is computationally infeasible to evaluate detailed models for multiple individual circuit elements. In this work we demonstrate a procedure for learning compact \JMHPBB{delayed} photocurrent models that are efficient enough to implement in large-scale circuit simulations, but remain faithful to the underlying physics. Our approach utilizes Dynamic Mode Decomposition (DMD), a system identification technique for learning reduced order discrete-time dynamical systems from time series data based on singular value decomposition. 
%
To obtain physics-aware device models, we simulate the excess carrier density induced by radiation pulses by solving \JMHPBB{numerically} the Ambipolar Diffusion Equation, then use the simulated internal state as training data for the DMD algorithm. Our results show that the significantly reduced order \JMHPBB{delayed} photocurrent models obtained via this method accurately approximate the dynamics of the internal excess carrier density \JMH{--} which can be used to calculate the induced current at the device boundaries \JMH{--} while remaining compact enough to incorporate into larger \JMHPBB{circuit} simulations.
\end{abstract}


\section{Introduction}\label{JMH:sec:intro}
Ionizing radiation can affect operation of \JMHREV{electronics} in multiple application contexts including space, terrestrial and manmade nuclear environments. Specifically, fluence of ionizing radiation generates electron-hole pairs within semiconductor devices, \JMHREV{resulting in excess currents flowing through the devices}. Such excess current is not present during device operation in a normal environment and its impact on electronic systems can be catastrophic, ranging from instantaneous interruptions in service, loss of stored memory, and even burnout of the entire system. For example spacecraft depend on electronic components that must perform reliably over missions measured in years and decades and space radiation is a primary source of degradation, reliability issues, and potentially failure for these electronic components. Physics-based modeling and simulation of radiation effects on electronic systems in various radiation environments can facilitate understanding of the mechanisms governing the radiation response of the electronic materials, parts, and systems, and can be used to devise ways to mitigate radiation effects, and \JMHPBB{create} new materials and devices that are \JMHREV{resilient} to radiation exposure. 

Thus, computational analysis of radiation effects on electronic systems has utility ranging from guiding the initial designs of systems, setting up the design of experiments, and final qualification. At a device level the excess carrier behavior can be accurately modeled by the Drift-Diffusion equations (DDE)\cite{JMH:Selberherr_84_BOOK} given by
\begin{align}\label{JMH:eq:DD1}
    \nabla \cdot (\epsilon \mathbf{E}) &= q (p - n + C) \\\label{JMH:eq:DD2}
    \frac{\partial n}{\partial t} &= \nabla \cdot (n \mu_n \mathbf{E} + D_n \nabla n) - R + g \\ \label{JMH:eq:DD3}
    \frac{\partial p}{\partial t} &= \nabla \cdot (p \mu_p \mathbf{E} + D_p \nabla p) - R + g
\end{align}
where $\JMHPBB{n}$ and $\JMHPBB{p}$ are the concentrations of the electrons and holes, respectively, $\mu_n$, $\mu_p$, $D_p$, and $D_n$ are the carrier mobilities and diffusivities, $C$ is the doping concentration, $\mathbf{E}$ is the electric field, $q$ is the electron charge, $R$ is the recombination rate, and $g$ is the generation rate.

Numerical solution of \eqref{JMH:eq:DD1}--\eqref{JMH:eq:DD3} using, e.g., a multi-dimensional finite element discretization of the device, forms the basis of the so-called Technology Computer-Aided Design (TCAD) simulators. However, \eqref{JMH:eq:DD1}--\eqref{JMH:eq:DD3} is a coupled system of nonlinear Partial Differential Equations (PDEs) whose accurate numerical solution can be very time consuming. As a result, although TCAD device simulators based on DDE could in principle be coupled to circuit simulators, their high computational cost makes the analysis of all but very small circuits computationally intractable. 

At the other end of the spectrum are the so-called compact device models, which are \JMHREV{computationally efficient} but rely on empirical approximations and/or simplified analytic solutions to the semiconductor transport equations. Often such models must be recalibrated for different operating regimes and inclusion of new physics may force redevelopment of the model from scratch. 

Data-driven techniques present an opportunity to automate the development of compact models by learning them directly from laboratory measurements \JMHPBB{and/or suitable synthetic data}. For such models to be practically useful though, they must be able to correctly predict the device response when the device is integrated into a circuit and exposed to a wide range of stimuli. The caveat is that in a laboratory settings one can only apply a limited type of signals and directly measure the device response, leading to ``sparse'' training sets. In contrast, traditional Machine Learning (ML) applications, such as natural language processing and image classification 
\cite{JMH:LeCun_15_Nature,JMH:Krizhevsky_17_CACM,JMH:He_15_arXiv} operate in ``big data'' environments. As a result, a compact device model learned solely from such ``sparse'' data may fail to generalize to all relevant analysis conditions because it will learn salient patterns in \JMHPBB{the} dataset rather than the causal physics underpinning the device operation. 

The latter, i.e., the fact that device behavior is governed by strict physics laws can be used to counter the lack of big data by incorporating the physics knowledge into the model development. Recent work on scientific Machine Learning \cite{JMH:Raissi_19_JCP,JMH:Rackauckas_20_arXiv,JMH:Bar-Sinai_19_PNAS} suggests this strategy is effective and can lead to generalizable models. 

The main goal of this work is to establish the viability of physics-aware machine learning for the development of compact data-driven photocurrent models that are efficient enough to allow large circuit simulations, while remaining faithful to the basic physics principles embodied by \JMHPBB{models such as} \eqref{JMH:eq:DD1}--\eqref{JMH:eq:DD3}. 
To that end we adopt a setting that \JMH{has} been used extensively in the past \JMHPBB{five} decades \JMHPBB{at Sandia National Laboratories} for the development of compact analytic photocurrent models. Although this setting uses a simplified version of the DDE equations, it provides a stepping stone towards future developments of data-driven models based on the fully coupled system \JMHPBB{\eqref{JMH:eq:DD1}--\eqref{JMH:eq:DD3}}. In so doing we are able to leverage a wealth of experiences and physics knowledge accumulated through the use of the existing compact models, as well as provide a reference point for evaluation of our data-driven models. 

We have organized the paper as follows. Section \ref{JMH:sec:ADE} reviews the current state-of-the art in compact photocurrent models and establishes the physics basis for our data-driven model. Section \ref{JMH:sec:FEM} briefly summarizes the finite element discretization of the physics model used to generate synthetic training data. The core of the paper is Section \ref{JMH:sec:DMD} where we use Dynamic Mode Decomposition ideas \cite{JMH:Schmid_10_JFM,JMH:Proctor_16_SIAM,JMH:Tu_14_JCD} to develop our model. Section \ref{JMH:sec:num} contains computational results highlighting the performance of the model as well as comparison with published results in \cite{JMH:Axness_04_JAP}. We summarize our findings and discuss future research in Section \ref{JMH:sec:concl}.


\section{Physics-Based Compact Analytic Photocurrent Models}\label{JMH:sec:ADE}

\begin{figure}[htbp] 
   \centering
   \includegraphics[width=0.35\linewidth]{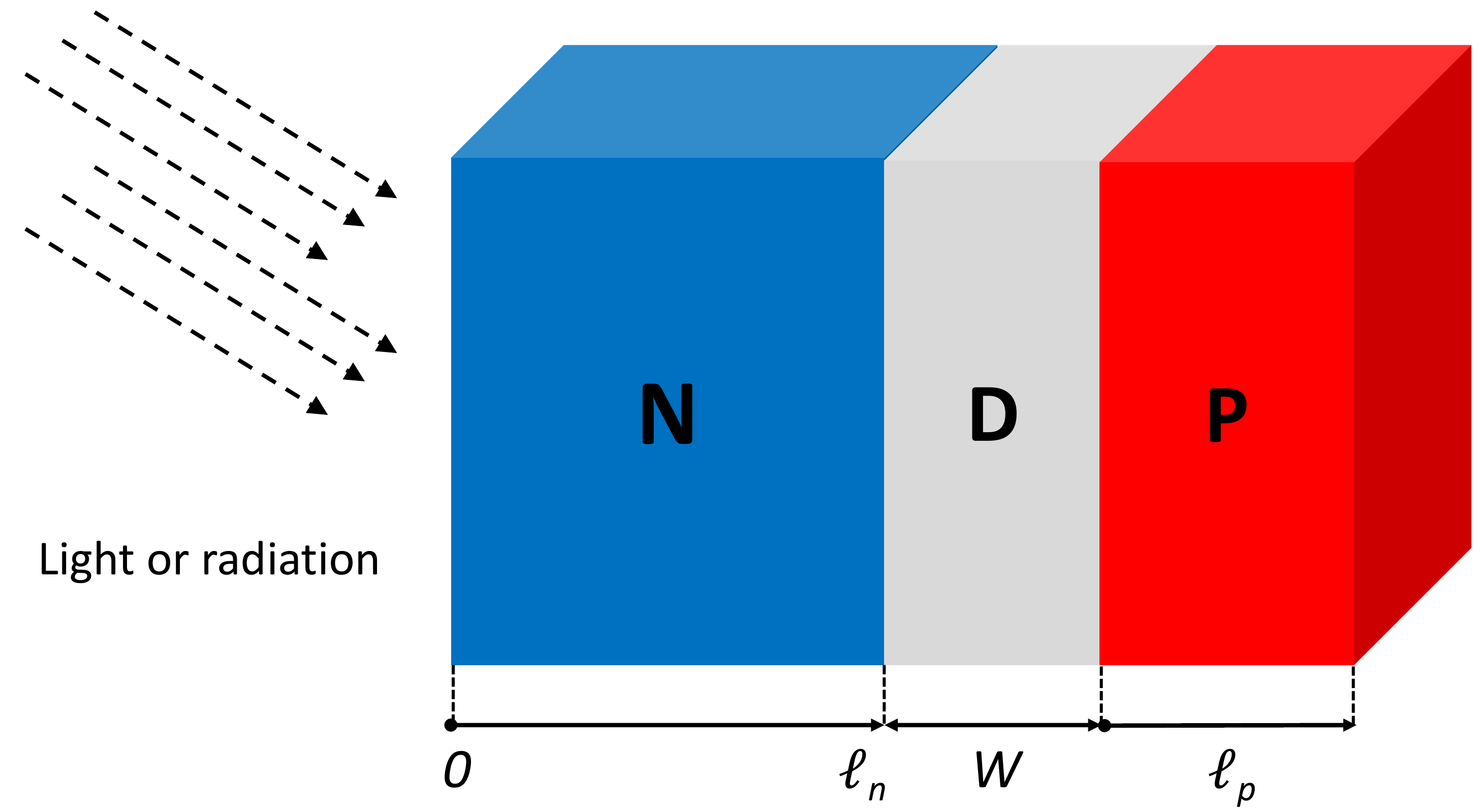} 
   \caption{\scriptsize Traditional compact photocurrent model development splits \JMH{each $PN$-junction within a} device into a depletion region D \JMHPBB{having width $W$}, and quasi-neutral $P$ and $N$ regions \JMHPBB{with lengths $\ell_n$ and $\ell_p$, respectively}; see, e.g., \cite{JMH:Axness_04_JAP}.}
   \label{JMH:fig:pn-diode}
\end{figure}

Although the \JMHPBB{DDE} system \eqref{JMH:eq:DD1}-\eqref{JMH:eq:DD3} accurately describes the behavior of the excess carriers, it does not lend itself to exact analytical solution. To enable analytic approximation of the governing equations, most photocurrent models in use today follow the same basic approach as in \JMHPBB{the classic paper} \cite{JMH:Wirth_64_IEEE_TNS} and split \JMH{each $PN$-junction within a} device into a depletion region and quasi-neutral $P$\JMH{-} and $N$-regions; see Fig.~\ref{JMH:fig:pn-diode}.
Carriers in the depletion region are quickly converted to photocurrent and yield the so-called prompt photocurrent $I_{\mathrm{prompt}}$. Carriers in the $P$ and $N$ regions have a delayed response and produce the \emph{delayed} photocurrents $I_p$ and $I_n$, respectively. As a result, the total photocurrent is given by
$$
I_{\mathrm{total}} = I_{\mathrm{prompt}} + I_p + I_n \,. 
$$
To calculate these currents one makes additional simplifying assumptions. \JMH{The first is the electrical neutrality approximation, which stipulates that the excess electron and hole densities are equal throughout the entire device. The second is the congruence assumption, which states that the electron and hole fluxes into or} out of any region must be equal; see, e.g., \cite{JMH:Kerr_12_SAND}. Under these assumptions the \JMHPBB{DDE} model \eqref{JMH:eq:DD1}-\eqref{JMH:eq:DD3} in the $P$ and $N$ regions can be replaced by the Ambipolar Diffusion Equation (ADE)
\begin{equation}\label{JMH:eq:ADE}
    \frac{\partial u}{\partial t} = D_a \nabla^2 u - \mu_a \mathbf{E} \cdot \nabla u - \frac{1}{\tau_a} u + g
\end{equation}
where $u$, $D_a$, $\mu_a$ and $\tau_a$ are the excess carrier density (electrons or holes), the ambipolar diffusion coefficient, the ambipolar mobility and the carrier lifetime, respectively. In general, these parameters may depend on the excess carrier density and, as a result, \eqref{JMH:eq:ADE} is still a nonlinear PDE. However, for moderate \JMH{radiation} dose rates these coefficients can be approximated by constant values and the ADE becomes a linear parabolic PDE. The final assumption is that the depletion region width $W$ is not affected by the excess carriers and is a constant. 
Under these assumptions, the prompt photocurrent is modeled as 
$$
I_{\mathrm{prompt}} = qgAW\,,
$$ 
where $A$ is the effective area of the \JMH{$PN$}-junction, while $I_p$ and $I_n$ are modeled by the ADE \eqref{JMH:eq:ADE}. 

At this point traditional compact photocurrent model development proceeds with deriving analytic approximations for the solutions of \eqref{JMH:eq:ADE} in one dimension and using them to obtain expressions for the delayed photocurrents. Early work \cite{JMH:Wirth_64_IEEE_TNS} considered unbounded $P$ and $N$ regions and negligible electric fields. The resulting Wirth-Rogers model tends to overestimate the photocurrent as it neglects the effects of an ohmic contact at a finite distance from the depletion region \cite{JMH:Axness_04_JAP}. Subsequent work \cite{JMH:Enlow_88_IEEE_TNS} relaxed these conditions, assumed that $\mathbf{E}$ is constant, and used approximate Laplace transforms to obtain analytic expressions for $I_p$ and $I_n$. However, approximation of the Laplace transform results in a model that yields unphysical current estimates when $\mathbf{E}$ exceeds roughly $\SI{10}{\volt\per\cm}$; see \cite{JMH:Wunsch_92_IEEE_TNS}. The latter work used Fourier analysis techniques to develop \JMHPBB{a more accurate}  photocurrent model that avoids these drawbacks. The model in \cite{JMH:Wunsch_92_IEEE_TNS} was further improved in \cite{JMH:Axness_04_JAP} by using a transformation of \eqref{JMH:eq:ADE} into an inhomogeneous heat equation and solving the latter \JMHPBB{exactly} by Fourier techniques. 
Another popular photocurrent model is the Fjeldly Model \cite{JMH:Fjeldly_01_IEEETNS}. Unlike the Axness-Kerr model \cite{JMH:Axness_04_JAP}, which solves the time-dependent equation \eqref{JMH:eq:ADE} exactly, the Fjeldly Model uses a steady-state solution of ADE combined with an RC delay circuit to achieve time dependence. 

\JMHPBB{In this work we adopt the above setting and focus on the development of physics-aware compact data-driven models for the delayed photocurrents in the $P$\JMH{-} and $N$-regions that can be used as "plug-and-play" substitutes for conventional models. The core idea is to replace the \emph{simplified analytic solutions} of the ADE, comprising the basis of most standard compact models, by a \emph{discrete-time dynamical system} approximating the flow map of the ADE, i.e. the ``solution operator'' that returns the internal device state for any given external input.} 

\JMHPBB{To that end we apply a Dynamic Mode Decomposition (DMD) \cite{JMH:Schmid_10_JFM} approach to samples of the internal state (carrier density) of the device obtained by solving \eqref{JMH:eq:ADE} numerically by a Finite Element Method (FEM). In this paper we assume that all device parameters such as diffusion coefficients, carrier lifetimes and doping concentrations are known. In this case ADE already contains all the necessary physics information and development of physics-aware compact data driven models can be done entirely from synthetic data\footnote{\JMHPBB{In this setting the resulting compact model can be interpreted as a discrete reduced order model of the flow map generated by the ADE.}}.}

\JMHPBB{A more general setting occurs when one or more of the ADE parameters are either unknown or have large uncertainties. In this case the development of the compact model must also include a parameter identification step to refine the ADE, which requires laboratory data. Since the purpose of this paper is to demonstrate the viability of DMD as an effective tool for the generation of compact device models, detailed discussion of this more general setting is beyond the scope of this paper and will be addressed in a forthcoming work.}

%

\JMH{Regardless} \JMHPBB{of the particular setting though the physics-awareness of our models stems from the fact that they represent \emph{approximations of the flow map engendered by the physics model}, and built from \emph{simulated internal states} of this model. 
These states contain physics information that cannot be obtained by laboratory instruments, which can typically only measure the currents at the device terminals. The latter may not be "rich enough" for a traditional ML approach, as well as for DMD, to obtain a reliable model of the underlying causal physics.}

\section{Numerical solution of the ADE} \label{JMH:sec:FEM}
We consider the ADE on the space-time domain $\Omega := \mathcal{X} \times \mathcal{T} \subset \mathbb{R}^2$, where $\mathcal{X} := (0,\ell)$ and $\mathcal{T} := (0,t_{\mathrm{final}})$. Without loss of generality, one may assume that $\mathcal{X}$ is the $N$-region of the device; see Fig.~\ref{JMH:fig:pn-diode} and so we set $\ell=\ell_n$. 
To obtain a well-posed problem, we augment equation \eqref{JMH:eq:ADE} with homogeneous initial and boundary conditions, i.e., 
$$
u(x,0) = 0\quad\forall x\in \mathcal{X} 
\quad\mbox{and}\quad
u(0,t) = u(\ell,t) = 0\quad \forall t \in \mathcal{T} \,.
$$
The homogenous initial condition corresponds to the fact that at $t=0$ there are no excess carriers present in the device. The boundary condition choice corresponds to assuming infinite carrier recombination velocities and ohmic contacts at $x=0$ and the boundary of the depletion region $x=\ell$; see \cite{JMH:Axness_04_JAP}. More general boundary conditions not requiring these assumptions can also be considered but are not necessary for the purpose of this work. 

We will simulate the internal state of the device by using the method of lines to solve the ADE numerically. For the spatial discretization we consider a standard Galerkin finite element method and then solve the resulting system of Ordinary Differential Equations (ODEs) using an implicit numerical integration scheme. For completeness we briefly review the discretization process below. 

Let $\mathcal{X}^h$ denote a uniform\footnote{Utilizing a variable mesh spacing with increased node density near the boundary may provide some potential advantages such as more accurate gradient estimation at the edge points, however in the context of this work we restrict our attention to uniform meshes.} partition of $\mathcal{X}$ into $n+1$ elements $\kappa_i$ with vertices $\{x_i\}_{i=0}^{n+1}$, i.e., $\kappa_i = [x_i,x_{i+1}]$, $i=0,\JMHPBB{\ldots},n$, $x_0 = 0$ and $x_{n+1} = \ell$. The mesh parameter is given by $h=\frac{\ell}{n+1}$. 


As usual, $L^2(\mathcal{X})$ denotes the space of all square integrable functions on $\mathcal{X}$ with norm and inner product denoted by $\|\cdot\|_0$ and $(\cdot,\cdot)_0$, respectively, and $H^1_0(\mathcal{X})$ is the Sobolev space of order one whose elements are constrained to vanish at the boundary points. The weak variational form of the ADE is then given by \emph{seek} $u\in H^1_0(\mathcal{X})$ \emph{such that}
\begin{equation}\label{JMH:eq:weakADE}
(u_t,v)_0 + Q(u,v) = (g,v)_0\quad\forall v \in H^1_0(\mathcal{X})\,.
\end{equation}
 The bilinear form $Q(\cdot,\cdot):H^1_0(\mathcal{X})\times H^1_0(\mathcal{X}) \mapsto \mathbb{R}$ is defined as 
 \begin{equation}\label{JMH:eq:Q}
 Q(u,v) = D_a\left(u_x,v_x\right)_0 + \mu_a \mathbf{E}\left(u_x,v\right)_0 + \frac{1}{\tau_a}\left(u,v\right)_0 \,.
 \end{equation}
To discretize \eqref{JMH:eq:weakADE} in space we consider a nodal (Lagrangian) conforming finite element subspace $V^h_0\subset H^1_0(\mathcal{X})$; see, e.g., \cite{JMH:Ciarlet_02_BOOK}. Let $\{v_i\}_{i=1,n}$ be the standard nodal basis having the property that $v_i(x_j) = \delta_{ij}$. We then seek an approximate solution of \eqref{JMH:eq:ADE} as
\begin{equation}\label{JMH:eq:uh}
u_h(x,t) = \sum_{i=1}^{n} u_i(t) v_i(x) \,,
\end{equation}
where $\mathbf{u}(t) = (u_1(t),\ldots,u_n(t)) \in \mathbb{R}^n$ is a vector of unknown solution coefficients. 
Inserting \eqref{JMH:eq:uh} into the weak form \eqref{JMH:eq:weakADE} and restricting the test space to $V^h_0$ then yields the system of ODEs
\begin{equation}\label{JMH:eq:ODE}
M \dot{\mathbf{u}}(t) + {K} \mathbf{u}(t) = \mathbf{g} 
\end{equation}
where $M,{K}\in\mathbb{R}^{n\times n}$ are the (consistent) finite element mass and stiffness matrices with elements
$$
M_{ij} = \left(v_i,v_j\right)_0
\quad\mbox{and}\quad
{K}_{ij} = Q(v_i,v_j)\,,
$$
respectively, and $\mathbf{g}(t)\in \mathbb{R}^n$ is a discrete source term with $g_i(t) = (g,v_i)_0$. 

The ODE system \eqref{JMH:eq:ODE} can be solved by any standard time-integration scheme. However, in general \eqref{JMH:eq:ODE} is stiff and an implicit scheme is preferred. In this paper, we use an implicit multi-step variable-order routine based on a backward differentitation formula (BDF) for approximating the state derivative; for more details see \cite{JMH:Shampine_97_SIAM}. This method is included by default in the \texttt{scipy.integrate} submodule within the SciPy v1.5.1 package for Python 3.


\section{A Dynamic Mode Decomposition Compact Photocurrent Model} \label{JMH:sec:DMD}

Assume that an approximate solution of the ODE \eqref{JMH:eq:ODE} is available at uniformly spaced time steps $t_k:=k\Delta t$, $k=0,\ldots,m$, for some sampling interval $\Delta t > 0$. Then, the approximate numerical solution of the ADE can be represented as a linear discrete-time dynamical system acting on samples of the state $\mathbf{u}$ \JMHPBB{and the input $\mathbf{g}$}, i.e.,
\begin{equation}
    \mathbf{u}_{k+1} = A \mathbf{u}_{k} + B \mathbf{g}_{k}
\end{equation}
where
\begin{align}
    \mathbf{u}_{k} &:= [u_1(k \Delta t) \cdots u_n(k \Delta t)]^\mathsf{T} \\
    \mathbf{g}_{k} &:= g(k \Delta t)
\end{align}
and $A, B \in \mathbb{R}^{n \times n}$ are linear maps. Expressing the system in this form facilitates the use of many familiar system identification techniques. In particular, Dynamic Mode Decomposition (DMD) is a data-driven method for learning the maps $A$ and $B$ from a time series of state and input measurements $\{\mathbf{u}_{k}, \mathbf{g}_{k}\}_{k=0}^m$, with the goal of identifying a small number of dominant dynamic modes. Below we summarize the DMD algorithm adapted for control inputs as described in \cite{JMH:Proctor_16_SIAM}. By organizing the samples into the following matrices
\begin{equation}
    X  = \begin{bmatrix}
        | & & | \\
        \mathbf{u}_{0} & \cdots & \mathbf{u}_{m-1} \\
        | & & |
    \end{bmatrix} \,; \quad
    X'  = \begin{bmatrix}
        | & & | \\
        \mathbf{u}_{1} & \cdots & \mathbf{u}_{m} \\
        | & & |
    \end{bmatrix} \,\JMH{;} \quad\mbox{and}\quad
    G = \begin{bmatrix}
        | & & | \\
        \mathbf{g}_{0} & \cdots & \mathbf{g}_{m-1} \\
        | & & |
    \end{bmatrix},
\end{equation}
we can express the linear relationships within the data as
\begin{equation}
    X' = AX + BG = \begin{bmatrix} A & B \end{bmatrix} \begin{bmatrix} X \\ G \end{bmatrix} =: \begin{bmatrix} A & B \end{bmatrix} S\JMH{.}
\end{equation}
Therefore the maps $A$ and $B$ can be approximated by
\begin{equation}\label{JMH:eq:DMD_equation}
    \begin{bmatrix} A & B \end{bmatrix} \approx \begin{bmatrix} \bar{A} & \bar{B} \end{bmatrix} := X' S^\dagger
\end{equation}
where $\dagger$ indicates the Moore-Penrose pseudoinverse. An efficient and accurate algorithm for estimating the pseudoinverse of a rectangular matrix is \JMH{realized via} truncated singular value decomposition. The matrix of samples $S$ can be factored as
\begin{equation}\label{JMH:eq:Omega_trun_SVD}
    S = U \Sigma V^\mathsf{T} = \begin{bmatrix} \tilde{U} & \tilde{U}_{\text{trun}} \end{bmatrix} \begin{bmatrix} \tilde{\Sigma} & 0 \\ 0 & \tilde{\Sigma}_{\text{trun}} \end{bmatrix} \begin{bmatrix} \tilde{V}^\mathsf{T} \\ \tilde{V}_{\text{trun}}^\mathsf{T} \end{bmatrix} \approx \tilde{U} \tilde{\Sigma} \tilde{V}^\mathsf{T}
\end{equation}
where $U \in \mathbb{R}^{n \times n}$, $\tilde{U} \in \mathbb{R}^{n \times p}$, $\Sigma \in \mathbb{R}^{n \times m}$, $\tilde{\Sigma} \in \mathbb{R}^{p \times p}$, $V^\mathsf{T} \in \mathbb{R}^{m \times m}$, and $\tilde{V}^\mathsf{T} \in \mathbb{R}^{p \times m}$. Here $_{\text{trun}}$ denotes the $m-p$ truncated singular values, so that the $p$ greatest singular values are kept. In practice, one sets an error tolerance and truncates all singular values below this threshold. Now the pseudoinverse of $S$ is naturally approximated by
\begin{equation}\label{JMH:eq:pseudo_inv_approx}
    S^\dagger = \tilde{V} \tilde{\Sigma}^{-1} \tilde{U}^\mathsf{T}
\end{equation}
Substituting equation \eqref{JMH:eq:pseudo_inv_approx} into \eqref{JMH:eq:DMD_equation} gives
\begin{align}
    A &\approx \bar{A} \approx X' \tilde{V} \tilde{\Sigma}^{-1} \tilde{U_1}^\mathsf{T} \in \mathbb{R}^{n \times n} \\
    B &\approx \bar{B} \approx X' \tilde{V} \tilde{\Sigma}^{-1} \tilde{U_2}^\mathsf{T} \in \mathbb{R}^{n \times n},
\end{align}
where $\tilde{U} = \begin{bmatrix} \tilde{U}_1^\mathsf{T} & \tilde{U}_2^\mathsf{T} \end{bmatrix}^\mathsf{T}$ with $\tilde{U}_1 \in \mathbb{R}^{n \times p}$, $\tilde{U}_2 \in \mathbb{R}^{n \times p}$.

We can achieve a more compact model by incorporating the projection $\tilde{\mathbf{u}} = P\mathbf{u}$ of the state onto the canonical dynamic mode coordinates. In the same manner as equation \eqref{JMH:eq:Omega_trun_SVD}, we factor the matrix $X' \approx \hat{U} \hat{\Sigma} \hat{V}^\mathsf{T}$, where the truncations are chosen to preserve the $r$ greatest singular values. Then the projection onto dynamic mode coordinates is given simply by $P = \hat{U}^\mathsf{T} \in \mathbb{R}^{r \times n}$, hence the desired reduced order model obtained via DMD is
\begin{align}\label{JMH:eq:DMD_model}
    \tilde{\mathbf{u}}_{k+1} &= \tilde{A}\tilde{\mathbf{u}}_{k} + \tilde{B}\mathbf{g}_{k} \\
    \tilde{\mathbf{u}}_{0} &= \hat{U}^\mathsf{T} \mathbf{u}_{0} \\
    \mathbf{u}_{k} &\approx \hat{U} \tilde{\mathbf{u}}_{k}
\end{align}
where
\begin{alignat}{3}
    \tilde{A} &:= \hat{U}^\mathsf{T} \bar{A} \hat{U} & &= \hat{U}^\mathsf{T} X' \tilde{V} \tilde{\Sigma}^{-1} \tilde{U_1}^\mathsf{T} \hat{U} & &\in \mathbb{R}^{r \times r} \\
    \tilde{B} &:= \hat{U}^\mathsf{T} \bar{B} & &= \hat{U}^\mathsf{T} X' \tilde{V} \tilde{\Sigma}^{-1} \tilde{U_2}^\mathsf{T} & &\in \mathbb{R}^{r \times n}.
\end{alignat}

The parameter $p$ represents the number of dynamic modes to fit to the data, which controls the model precision. The parameter $r$ represents the number of modes to project onto, that is, the order of the final reduced-order model, which controls the model compactness. The case where $r > p$ usually results in diminished performance; $r = p$ retains exactly the same number of modes fit to the data in the compactified model; $r < p$ results in a more compact model, but ignores the $p - r$ least significant modes fit to the data, which can result in slightly reduced accuracy. In this work we will use $r = p$.

\subsection{Training of the compact model}\label{JMH:sec:train}
\JMHPBB{Construction of the DMD model \eqref{JMH:eq:DMD_model} requires training samples representing time series of state and input measurements $\{\mathbf{u}_{k}, \mathbf{g}_{k}\}_{k=0}^m$. To generate such samples we recall the assumption in Section \ref{JMH:sec:ADE} that all device parameters are known. In particular, here we consider a generic} \JMH{$PN$-junction device} \JMHPBB{characterized by the parameters in Table \ref{JMH:tab:ADEpars}. The values in this table are adapted from \cite{JMH:Axness_04_JAP} to enable a direct comparison with a published compact photocurrent model.}

\begin{table}[ht]
\caption{ADE parameters for a generic $PN$-junction device}
\centering
\scriptsize
\begin{tabular}{c l l l}
\hline
Parameter & Value & Units & Description  \\ [0.5ex] 
\hline
$\ell$       & $3.075 \times 10^{-2}$ & $\SI{}{\cm}$                      & $N$-region length \\
$D_a$        & $1.19 \times 10^1$     & $\SI{}{\cm\squared\per\second}$   & diffusion coefficient \\
$\mu_a$      & $4.64 \times 10^2$     & $\SI{}{\cm\squared\per\volt\per\second}$ & typical hole mobility for $Si$ \\
$\ell_n$        & $1.54 \times 10^{-2}$  & $\SI{}{\cm\squared\per\second}$   & diffusion length: $\sqrt{D_a\tau_a}$ \\
$\tau_a$     & $1.97 \times 10^{-5}$  & $\SI{}{\second}$                  & typical hole lifetime \\
$\mathbf{E}$ & $-20$ or $0$           & $\SI{}{\volt\per\cm}$             & electric field \\
$C$          & $1 \times 10^{17}$     & $\SI{}{\per\cm\cubed}$            & doping concentration \\
\JMHREV{$\widehat{g}$}     & $4.3\times 10^{22}$   & $\SI{}{\per\cm\cubed\per\second}$ & \JMHREV{max generation density}  \\
$u(x,t)$     & variable               & $\SI{}{\per\cm\cubed}$            & excess carrier concentration (ADE solution)
\\ [1ex]
\hline
\end{tabular}
\label{JMH:tab:ADEpars}
\end{table}

\JMHPBB{We then select a suitable set of generation density functions $\{g^k_{\mathrm{train}}\}_{k=1}^{M}$ and use the computational scheme in Section \ref{JMH:sec:FEM} to solve \eqref{JMH:eq:ADE} numerically with homogeneous initial and Dirichlet boundary conditions. Selection of the inputs $g^k_{\mathrm{train}}$ depends on the type of the anticipated testing input(s) $g_{\mathrm{test}}$ for the model and will be revisited in Section \ref{JMH:sec:num}.} 

Obtaining a numerical solution for the ADE requires proper scaling and non-dimensionalization of the governing equations. For convenience we scale the computational domain $\mathcal{X}$ so that the length of the $N$-region, i.e., $\ell$, becomes a unit of length and $t_{\mathrm{final}}$ becomes a unit of time. After rescaling the domain and the equations the computational domain $\mathcal{X}$ becomes the unit square, and the non-dimensional ADE coefficients are given by
\begin{equation}\label{JMH:eq:scaledADE}
\begin{array}{l}
D_a = 0.063 \\
\mu_a = 0.063 
\end{array}
;\quad
\begin{array}{l}
L_p    = 0.5 \\
\tau_a = 3.92 \\
\end{array}
\JMH{;}\quad\mbox{and}\quad
\begin{array}{l}
\mathbf{E} = -23.84 \ \mbox{or} \ 0 \\
\JMHREV{\widehat{g}} = 2.15 
\end{array} \,.
\end{equation}

We highlight that the \JMHPBB{numerical solution of the ADE} constitutes the "physics-based" element of our procedure. Specifically, the physics information is incorporated by using trusted \textit{a priori} dynamics models -- which are calibrated or driven by experimental measurements -- to generate training data from simulating the unobservable internal state of a device using robust numerical techniques. 

\begin{remark}
In this work we use the ADE as the physics basis for the data-driven model because it has been used to develop almost all compact photocurrent models in use today, i.e., it is an example of model that is trusted based on decades-long practical experiences. 
However, we emphasize that the DMD algorithm is also suitable for more complex physics-based models, such as the full drift-diffusion equations or detailed molecular dynamics simulations, with the capability of producing dramatically reduced order approximations that are feasible to implement in high-level circuit simulators but remain faithful to the underlying physics.
\end{remark}

\section{Simulation Results}\label{JMH:sec:num}
%
Assuming that a DMD model \eqref{JMH:eq:DMD_model} has been trained according to the procedure in Section \ref{JMH:sec:train}, we test its performance as follows. 
Let $g(x,t)$ be a target generation density for which we seek the response of our device. We sample $g(x,t)$ in space using the vertices defining the finite element mesh $\mathcal{X}^h$, and in time using a desired time step $\Delta t$ for a total of $m$ time steps.
This sampling produces the inputs $\mathbf{g}_k$ to the DMD model. We then set $\mathbf{u}_{0}=\mathbf{0}$ and use \eqref{JMH:eq:DMD_model} to recover the internal state of the device, i.e., the excess carrier concentration, at the mesh nodes $\{x_i\}_{i=1}^{n}$ for every $t_k = k\Delta t$:
$$
\tilde{\mathbf{u}}_{k+1} = \tilde{A}\tilde{\mathbf{u}}_{k} + \tilde{B}\mathbf{g}_{k} \,,
\quad k=0,\ldots,m-1\,.
$$
Each vector $\tilde{\mathbf{u}}_{k}$ induces a $C^0$ finite element function 
\begin{equation}\label{JMH:eq:DMD_density}
u_h^\mathrm{DMD}(x,t_k) = \sum_{i=1}^{n} \tilde{{u}}_i(t_k) v_i(x)
\end{equation}
which is the predicted internal carrier density. Using $u_h^\mathrm{DMD}$ and \JMHPBB{taking into account the homogeneous Dirichlet boundary condition $u_h^\mathrm{DMD}(x_0,t_k)=u_h^\mathrm{DMD}(x_{n+1},t_k) = 0$}, we define the approximate \JMHPBB{boundary photocurrent} at $t=t_k$ as 
\begin{equation}\label{JMH:eq:DMD_flux}
J_h^\mathrm{DMD}(x,t_k) = D_a \partial_x u_h^\mathrm{DMD}(x,t_k) = D_a \sum_{i=1}^{n} \tilde{u}_i(t_k) \partial_x v_i(x),
\end{equation}
\JMHPBB{for $x=x_0$ and $x=x_{n+1}$.} Note that owing to the local support of the basis functions $v_i(x)$, $J_h^\mathrm{DMD}(x_0,t_k)$ and $J_h^\mathrm{DMD}(x_0,t_k)$ only include contributions from the basis functions supported on elements $\kappa_0$ and $\kappa_n$, respectively. 
We then compare the predicted DMD flux with simulated experimental measurements of the current out of the device terminals. These measurements are obtained using the same numerical procedure as in Section \ref{JMH:sec:FEM}, \JMHPBB{i.e., by computing a finite element solution $u_h^\mathrm{FEM}(x,t_k)$ and setting}
\JMH{\begin{align}
	J_h^\mathrm{FEM}(x_0,t_k) &= D_a \partial_x u_h^\mathrm{FEM}(x_0,t_k) \label{JMH:eq:FEM_flux_left} \\
	J_h^\mathrm{FEM}(x_{n+1},t_k) &= D_a \partial_x u_h^\mathrm{FEM}(x_{n+1},t_k). \label{JMH:eq:FEM_flux_right}
\end{align}}

\subsection{Manufactured solution test}\label{JMH:sec:num-mnf}
%
We first evaluate the ability of the DMD model to reproduce an artificially manufactured solution:
\begin{equation}\label{JMH:eq:MNF_sol}
	u^{\mathrm{MNF}}(x,t) := t e^{-2t} \sin\big(\frac{3 \pi x}{\ell}\big).
\end{equation}
Observe that this solution satisfies homogenous initial and boundary conditions. Since \eqref{JMH:eq:MNF_sol} is expressed in closed-form, we can directly compute the boundary photocurrent of the manufactured solution:
\begin{equation}\label{JMH:eq:MNF_flux}
	J^\mathrm{MNF}(x,t) = D_a \partial_x u_h^\mathrm{MNF}(x,t) = D_a t e^{-2t} \big(\frac{3 \pi}{\ell}\big) \cos\big(\frac{3 \pi x}{\ell}\big).
\end{equation}
\JMHPBB{Thus, for this test we will compare $J^\mathrm{DMD}$ with the known manufactured solution current $J^\mathrm{MNF}$, rather than with the simulated current $J^\mathrm{FEM}$ in \eqref{JMH:eq:FEM_flux_left}--\eqref{JMH:eq:FEM_flux_right}.}
Substitution of $u^{\mathrm{MNF}}(x,t)$ into the governing equation \eqref{JMH:eq:ADE} yields a generation density
\begin{align}\label{JMH:eq:MNF_input}
	g^{\mathrm{MNF}}(x,t) &= \frac{\partial u^{\mathrm{MNF}}}{\partial t}(x,t) - D_a \frac{\partial^2 u^{\mathrm{MNF}}}{\partial x^2}(x,t) + \mu_a \mathbf{E}(x) \frac{\partial u^{\mathrm{MNF}}}{\partial x}(x,t) + \frac{1}{\tau_a} u^{\mathrm{MNF}}(x,t) \\
	&= (1 - 2t) e^{-2t} \sin\big(\frac{3 \pi x}{\ell}\big) + D_a t e^{-2t} \big(\frac{3 \pi}{\ell}\big)^2 \sin\big(\frac{3 \pi x}{\ell}\big) \\
	&\quad + \mu_a E(x) t e^{-2t} \big(\frac{3 \pi}{\ell}\big) \cos\big(\frac{3 \pi x}{\ell}\big) + \frac{1}{\tau_a} t e^{-2t} \sin\big(\frac{3 \pi x}{\ell}\big)
\end{align}
such that when the ADE \eqref{JMH:eq:ADE} is driven by $g^{\mathrm{MNF}}(x,t)$, \JMHPBB{its solution will exactly match} the desired manufactured solution $u^{\mathrm{MNF}}(x,t)$. The inputs $\mathbf{g}_k$ to the DMD model are obtained by sampling $g^{\mathrm{MNF}}(x,t)$ according to the method described earlier.

\begin{figure}[htbp] 
   \centering
   \textbf{Training input spatial envelopes}\par\medskip
   \adjincludegraphics[width=0.48\textwidth, Clip={.0\width} {.0\height} {.0\width} {.1\height}]{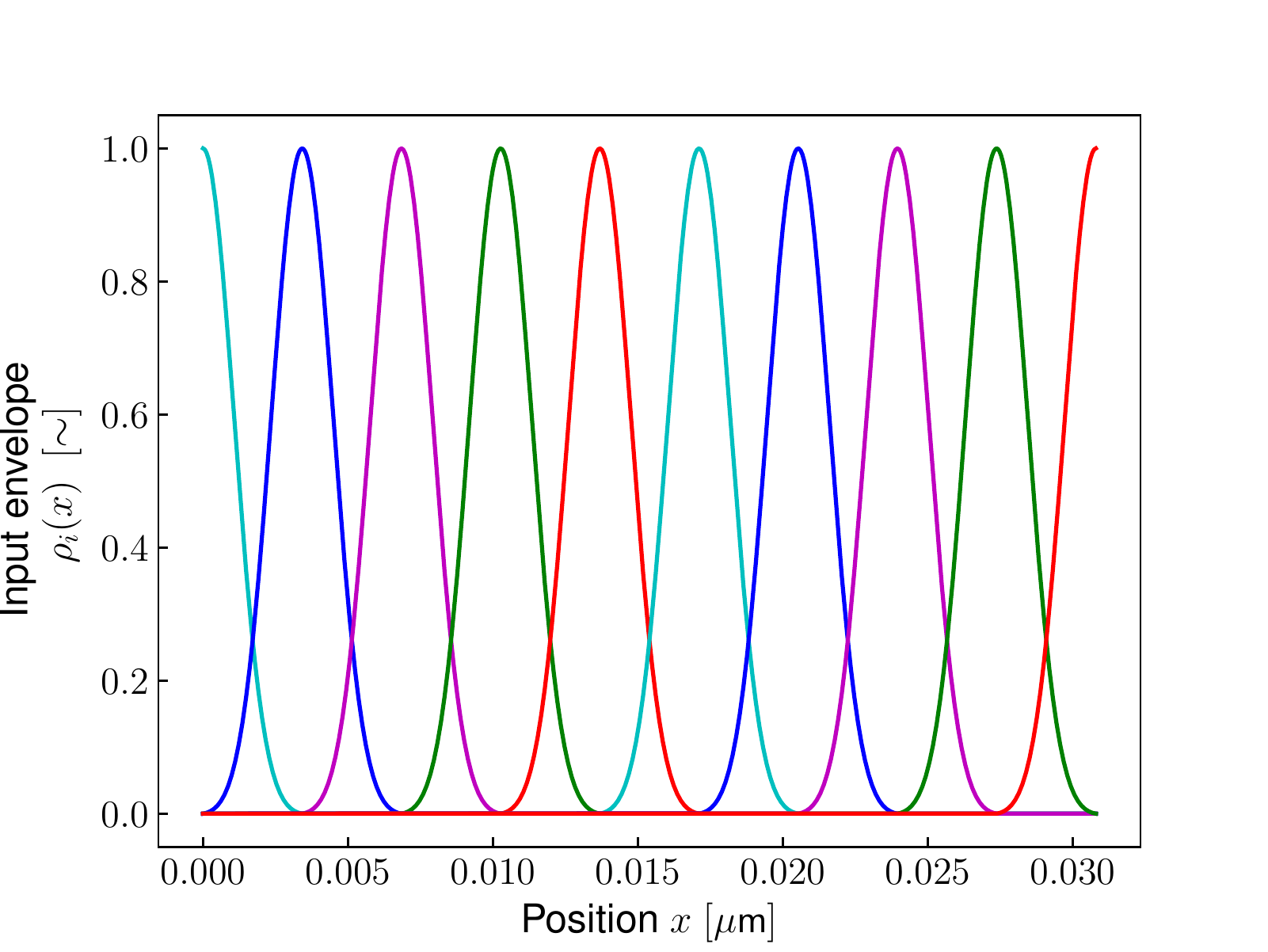}
   \caption{\scriptsize Generation density function $g_{\mathrm{train}}$ used to obtain the training set, and $g_{\mathrm{test}}$ used to verify the model performance. The square edges in the training input are included to excite a wider range of dynamic modes, which is due to the high-bandwidth content in the sharp transitions.}
   \label{JMH:fig:input_profiles}
\end{figure}

Since the input $g^{\mathrm{MNF}}(x,t)$ corresponding to the manufactured solution is spatially irregular, a spatially uniform training input will generally result in poor performance. To address this, we design a sequence of localized pulses which will excite different regions of the device. This sequence will then be used as the training input for the DMD model. We choose to use a Gaussian profile that has been windowed by a cosine function as the spatial envelope for the input pulses, where the window function is applied to restrict the support of the envelope to a compact interval. This profile is consistent with experimentally viable radiation doses; other reasonable choices include \textit{Lorenz} or \textit{Voigt} profiles, which reflect different radiation broadening mechanisms. The windowed Gaussian profile with center $x_i$ and support $[x_i-\frac{w}{2},x_i + \frac{w}{2}]$ is given by
\begin{equation}
	\rho_i(x) := \begin{cases}
	\cos\big(\pi \frac{x - x_i}{w}\big) \exp\big(-16\big(\frac{x - x_i}{w}\big)^2\big) & \text{if}\ x_i - \frac{w}{2} \leq x \leq x_i + \frac{w}{2} \\
	0 & \text{if}\ x < x_i - \frac{w}{2}\ \text{or}\ x > x_i + \frac{w}{2}
	\end{cases}
\end{equation}
where $x_i = i\frac{\ell}{N_{\mathrm{pulses}} - 1}$ for $i = 0,\dots,N _{\mathrm{pulses}}-1$, and $w = \frac{\ell}{N_{\mathrm{pulses}} - 2}$. The profiles $\rho_i(x)$ are illustrated in Figure \ref{JMH:fig:input_profiles}. Combining the spatial envelopes defined above with a square temporal envelope gives
\begin{equation}
g_i(x,t) =
\begin{cases}
\widehat{g} \rho_i(x) & \mbox{if} \ 0 \leq t \leq \SI{0.5}{\micro\second} \\[1ex]
0       & \mbox{otherwise}
\end{cases},
\end{equation}
\JMHPBB{where the value of $\widehat{g}$ is defined in \eqref{JMH:eq:scaledADE}.}
Now we can express the training input as
\begin{equation}\label{JMH:eq:mnf_train_input}
	g_{\mathrm{train}} = \sum_{i=0}^{N_{\mathrm{pulses}} - 1} g_i(x,t - t_i)
\end{equation}
where $t_i = i(\SI{5.0}{\micro\second})$ for $i = 0,\dots,N_{\mathrm{pulses}}-1$. For the manufactured solution \eqref{JMH:eq:MNF_sol}, we choose $N_{\mathrm{pulses}} = 10$.

\JMH{To generate the training samples, we solve \eqref{JMH:eq:ADE} with the source \eqref{JMH:eq:mnf_train_input} and the parameters \eqref{JMH:eq:scaledADE} on a mesh $\mathcal{X}^h$ comprising 512 uniform elements, and sample the solution in time at $\Delta t = \SI{0.005}{\micro\second}$ increments.}

Using the the manufactured photocurrent \eqref{JMH:eq:MNF_flux} and carrier density \eqref{JMH:eq:MNF_sol} and approximate \JMHREV{photocurrent} \eqref{JMH:eq:DMD_flux} and approximate carrier density \eqref{JMH:eq:DMD_density} derived from the DMD model, we define the following error quantities:
\begin{align}
    E_J(t) &:= | J^\mathrm{DMD}_h(t) - J^\mathrm{MNF}(t) | \label{JMH:eq:DMD_flux_error_MNF} \\
    E_u(t) &:= | u^\mathrm{DMD}_h(t) - u^\mathrm{MNF}(t) | \label{JMH:eq:DMD_density_error_MNF}
\end{align}
which will characterize the ability of the DMD model to reproduce manually selected dynamic modes. For simplicity, in this work the electric field is not considered an input to the system, so the DMD model must be trained separately for each unique electric field strength.

\begin{figure}
    \centering
    \begin{minipage}{0.48\textwidth}
        \centering
        \textbf{Manufactured \JMH{s}olution \JMH{f}lux \\ (no electric field)}\par\medskip
        \includegraphics[width=\textwidth]{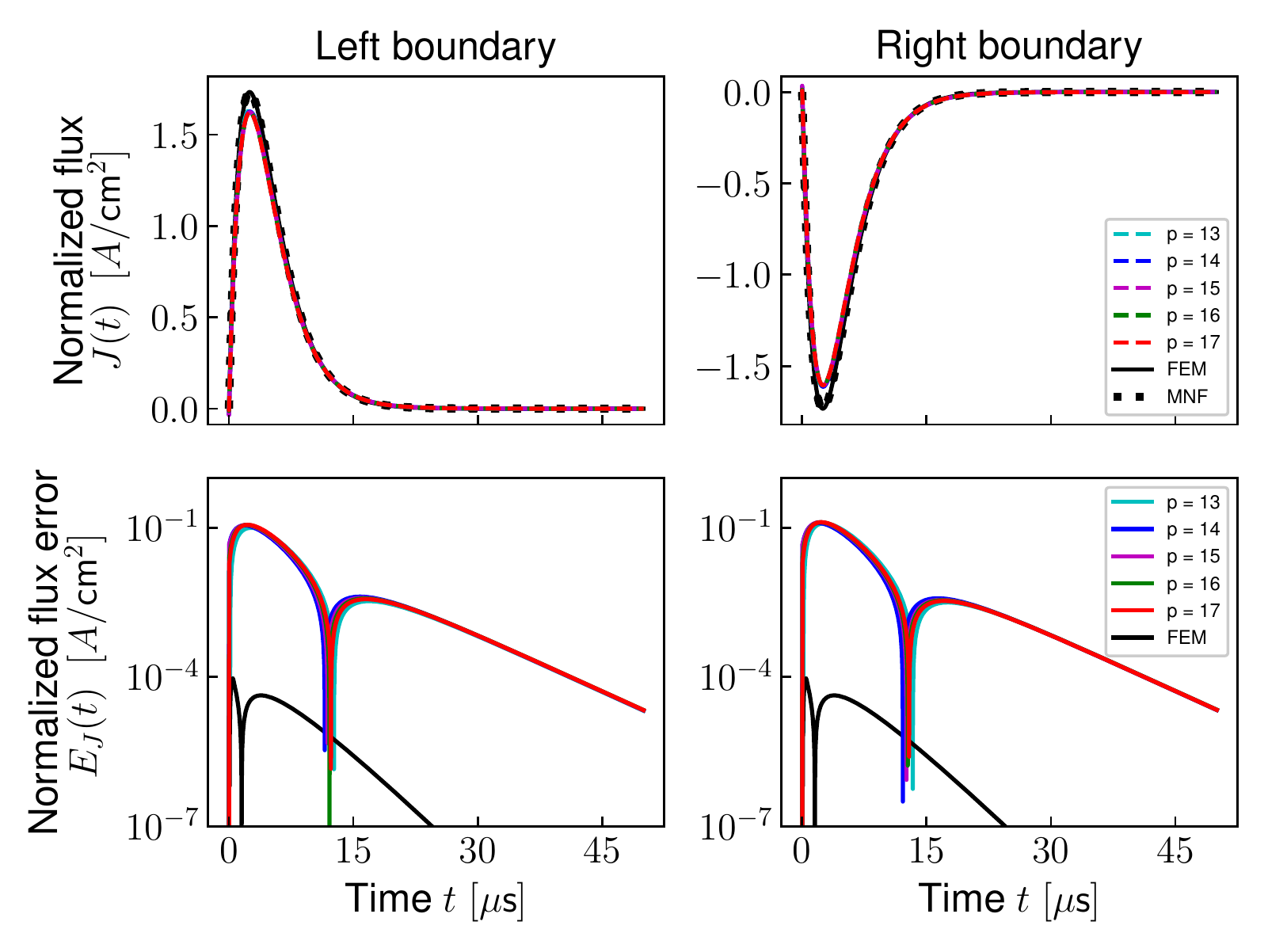}
        \caption{\scriptsize Normalized \JMHREV{photocurrent} due to the manufactured input \eqref{JMH:eq:MNF_input} with no electric field applied. The top two plots show the manufactured solution \eqref{JMH:eq:MNF_flux}, FEM solution \JMH{\eqref{JMH:eq:FEM_flux_left}-\eqref{JMH:eq:FEM_flux_right}}, and DMD solution \eqref{JMH:eq:DMD_flux}, and the bottom two plots show the DMD and FEM error \eqref{JMH:eq:DMD_flux_error_MNF}. The left two plots correspond to the flux out of the left side of the $N$-region, and similarly on the right.}
        \label{JMH:fig:mnf_flux_no_E}
    \end{minipage}\hfill
    \begin{minipage}{0.48\textwidth}
        \centering
        \textbf{Manufactured \JMH{s}olution \JMH{f}lux \\ (with electric field)}\par\medskip
        \includegraphics[width=\textwidth]{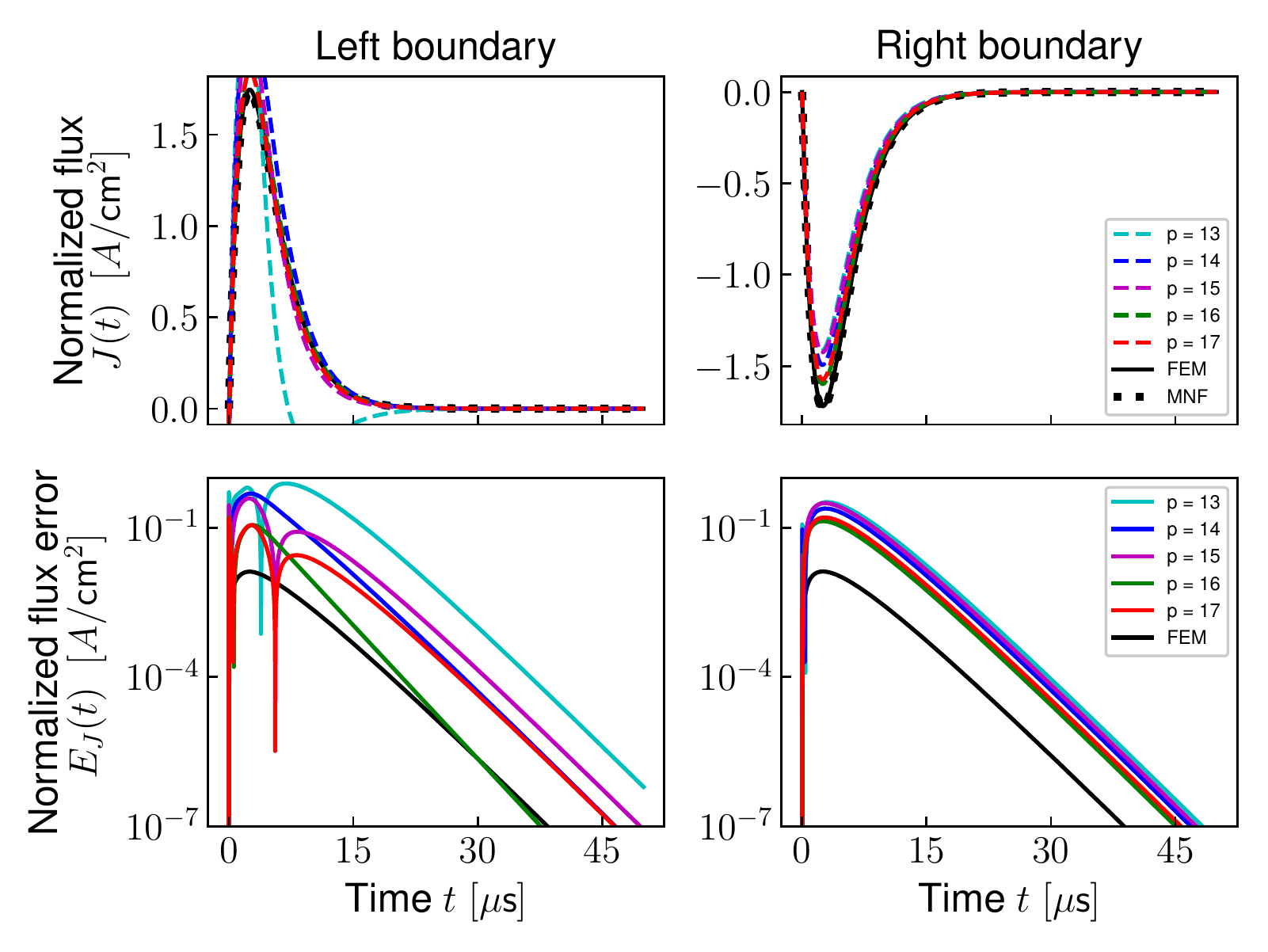}
        \caption{\scriptsize Normalized \JMHREV{photocurrent} due to the manufactured input \eqref{JMH:eq:MNF_input} with an electric field \eqref{JMH:eq:scaledADE} applied. The top two plots show the manufactured solution \eqref{JMH:eq:MNF_flux}, FEM solution \JMH{\eqref{JMH:eq:FEM_flux_left}-\eqref{JMH:eq:FEM_flux_right}}, and DMD solution \eqref{JMH:eq:DMD_flux}, and the bottom two plots show the DMD and FEM error \eqref{JMH:eq:DMD_flux_error_MNF}. The left two plots correspond to the flux out of the left side of the $N$-region, and similarly on the right.}
        \label{JMH:fig:mnf_flux_with_E}
    \end{minipage}
\end{figure}

\begin{figure}
    \centering
    \begin{minipage}{0.48\textwidth}
        \centering
        \textbf{Manufactured \JMH{s}olution \JMH{d}ensity (no electric field)}\par\medskip
        \includegraphics[width=\textwidth]{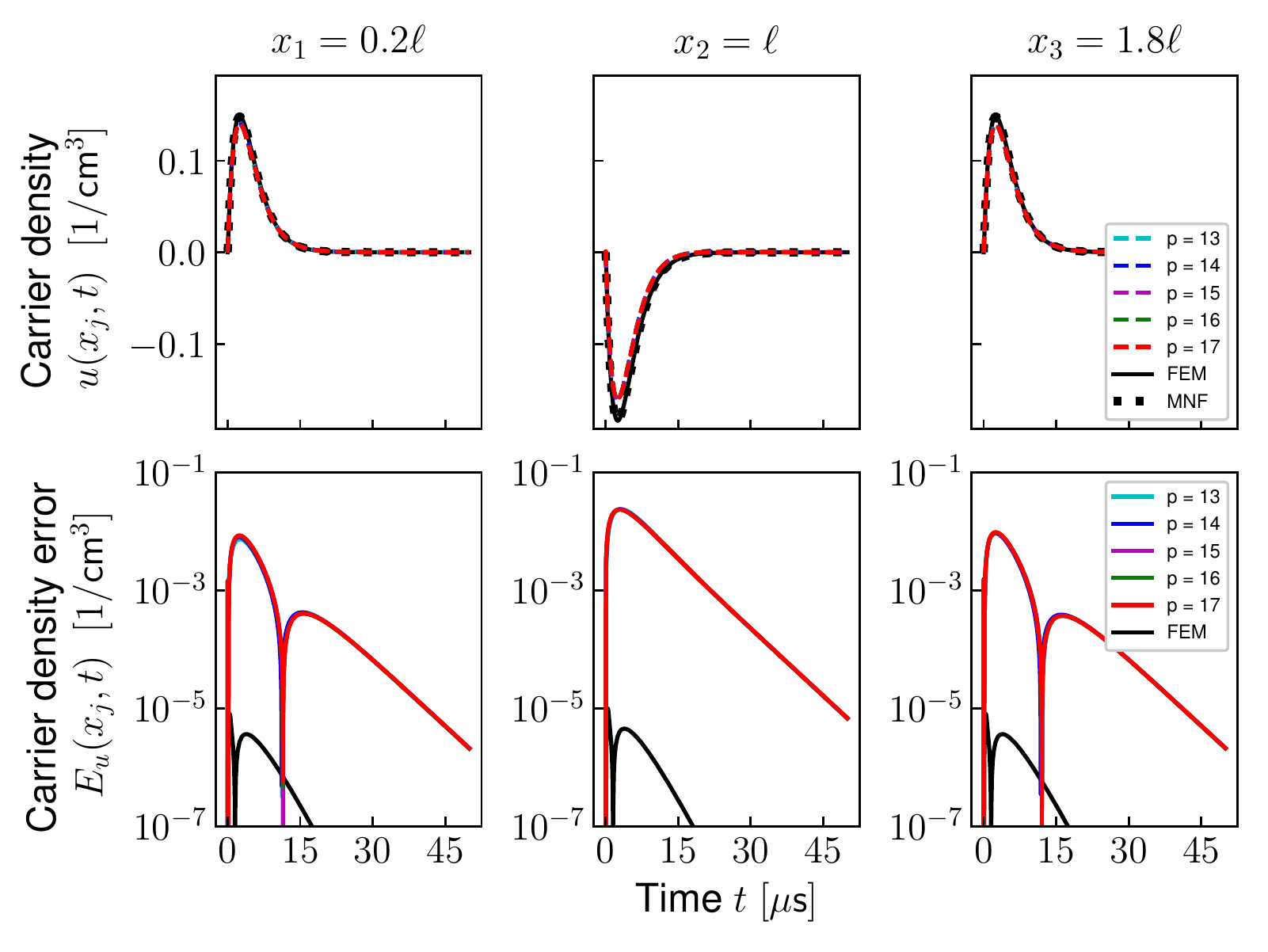}
        \caption{\scriptsize Simulated excess carrier density due to the manufactured input \eqref{JMH:eq:MNF_input} with no electric field applied. The top three plots show the manufactured solution \eqref{JMH:eq:MNF_sol}, FEM solution \eqref{JMH:eq:uh}, and DMD solution \eqref{JMH:eq:DMD_density}, and the bottom three plots show the DMD and FEM error \eqref{JMH:eq:DMD_density_error_MNF}.}
        \label{JMH:fig:mnf_density_no_E}
    \end{minipage}\hfill
    \begin{minipage}{0.48\textwidth}
        \centering
        \textbf{Manufactured \JMH{s}olution \JMH{d}ensity (with electric field)}\par\medskip
        \includegraphics[width=\textwidth]{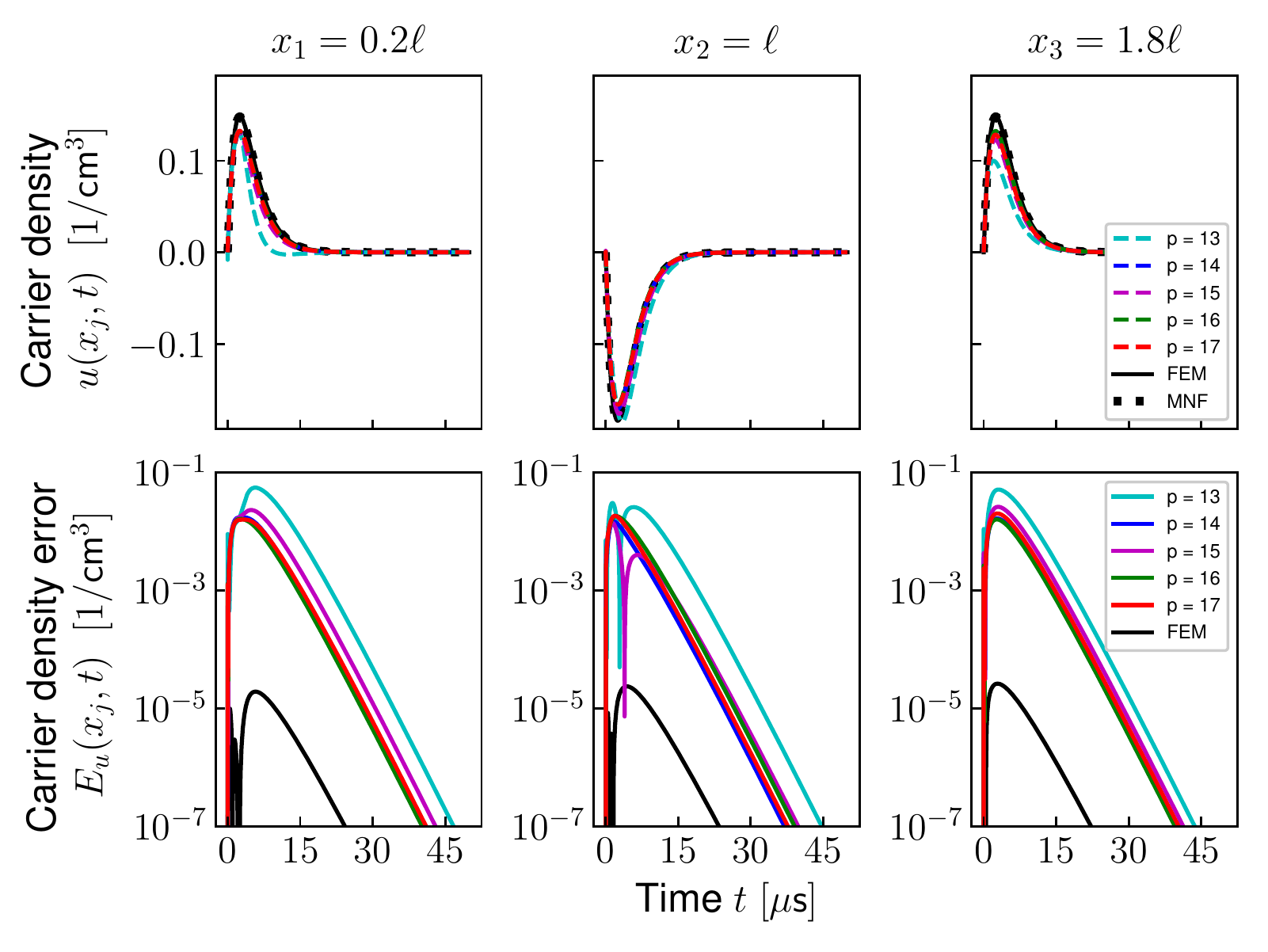}
        \caption{\scriptsize Simulated excess carrier density due to the manufactured input \eqref{JMH:eq:MNF_input} with an electric field \eqref{JMH:eq:scaledADE} applied. The top three plots show the manufactured solution \eqref{JMH:eq:MNF_flux}, FEM solution \eqref{JMH:eq:uh}, and DMD solution \eqref{JMH:eq:DMD_density}, and the bottom three plots show the DMD and FEM error \eqref{JMH:eq:DMD_density_error_MNF}.}
        \label{JMH:fig:mnf_density_with_E}
    \end{minipage}
\end{figure}

\begin{figure}
    \centering
    \begin{minipage}{0.48\textwidth}
        \centering
        \textbf{Manufactured \JMH{solution} \JMH{d}ensity \JMH{s}napshot (no electric field)}\par\medskip
        \includegraphics[width=\textwidth]{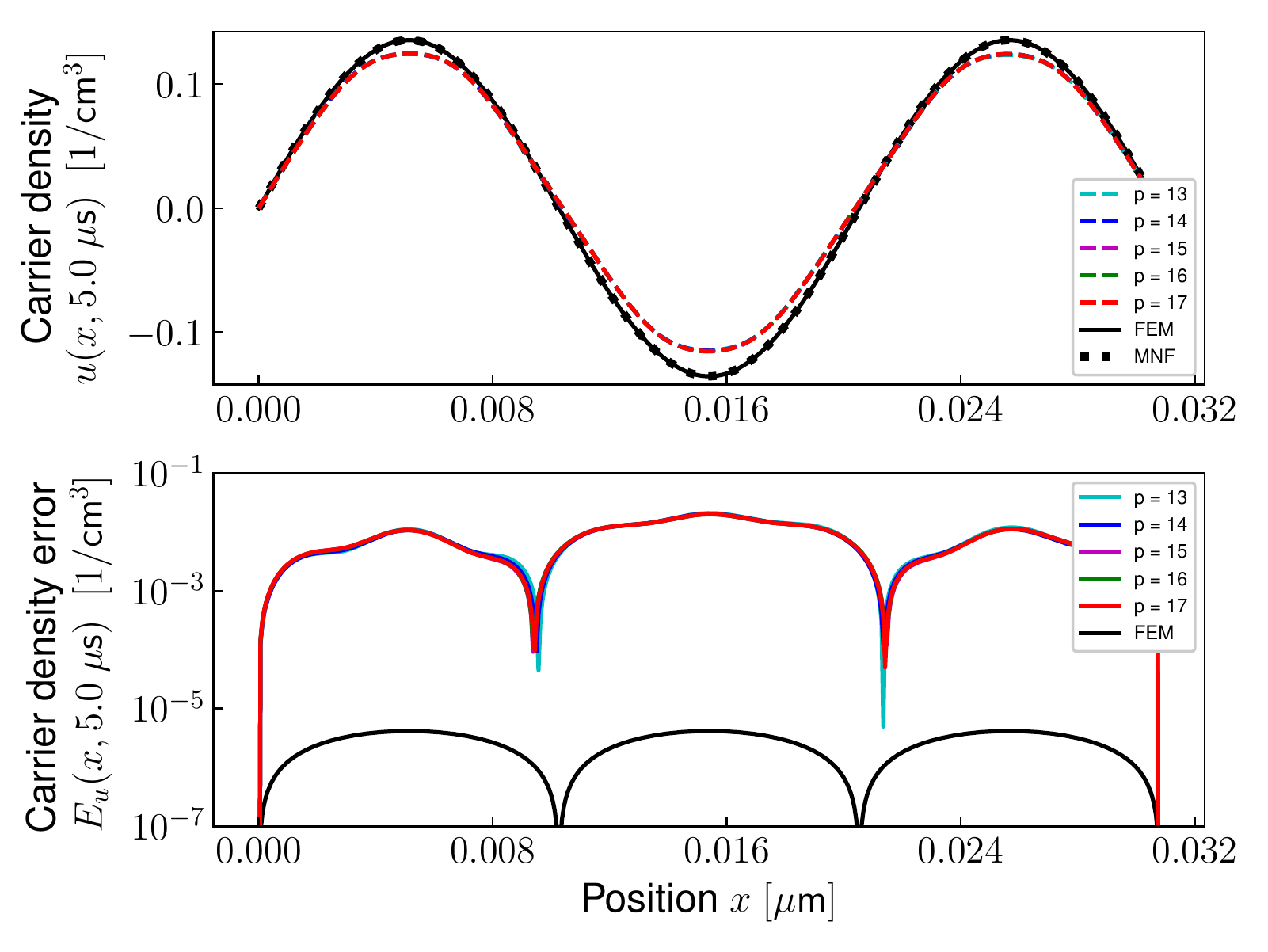}
        \caption{\scriptsize Snapshot of the manufactured excess carrier density at $t = \SI{5.0}{\micro\second}$ due to the manufactured input \eqref{JMH:eq:MNF_input} with no electric field applied. The top plot shows the manufactured solution \eqref{JMH:eq:MNF_flux}, FEM solution \eqref{JMH:eq:uh}, and DMD solution \eqref{JMH:eq:DMD_density}, and the bottom plot shows the DMD and FEM error \eqref{JMH:eq:DMD_density_error_MNF}.}
        \label{JMH:fig:mnf_snapshot_no_E}
    \end{minipage}\hfill
    \begin{minipage}{0.48\textwidth}
        \centering
        \textbf{Manufactured \JMH{solution} \JMH{d}ensity \JMH{s}napshot (with electric field)}\par\medskip
        \includegraphics[width=\textwidth]{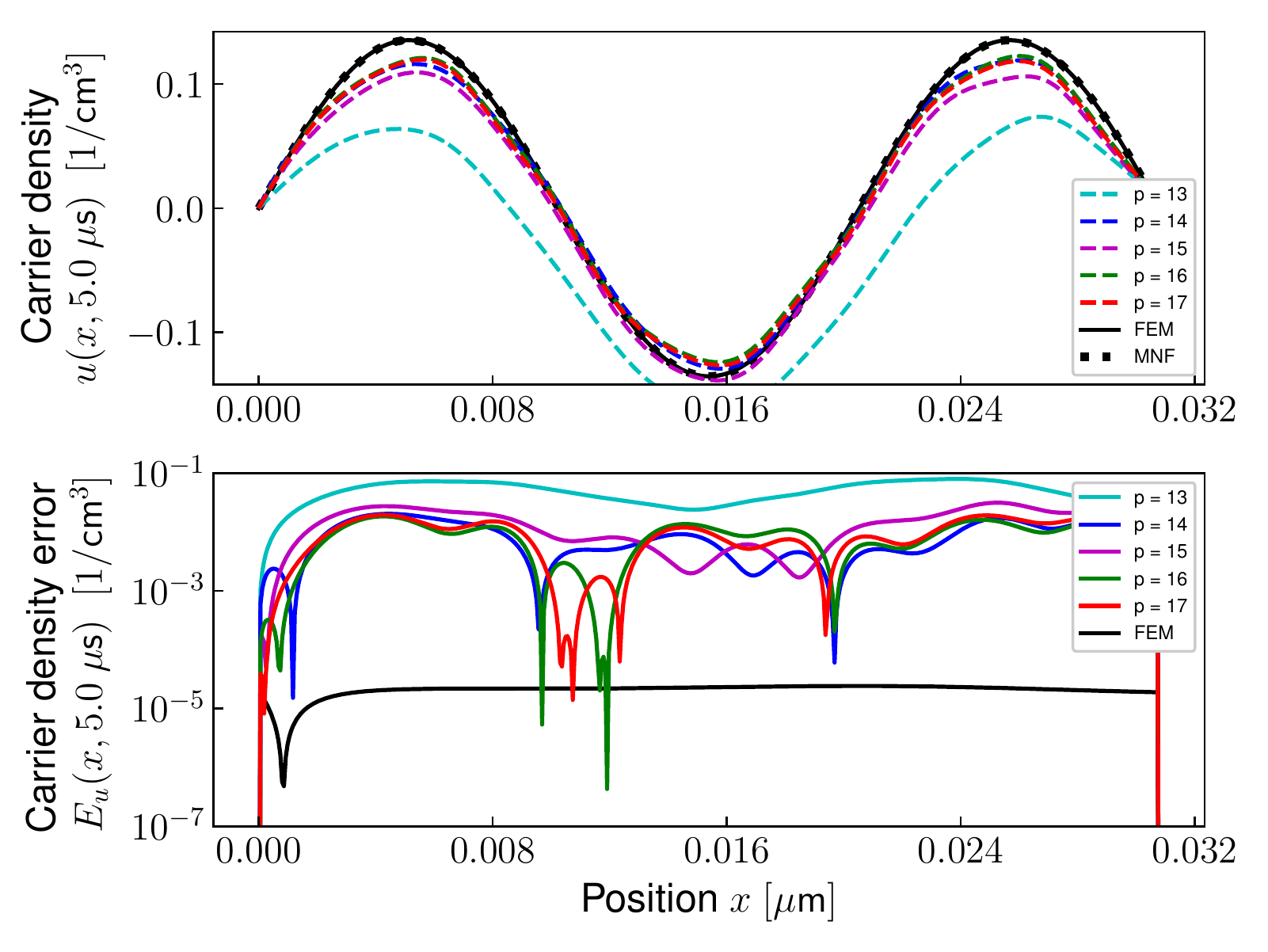}
        \caption{\scriptsize Snapshot of the manufactured excess carrier density at $t = \SI{5.0}{\micro\second}$ due to the manufactured input \eqref{JMH:eq:MNF_input} with an electric field \eqref{JMH:eq:scaledADE} applied. The top plot shows the manufactured solution \eqref{JMH:eq:MNF_flux}, FEM solution \eqref{JMH:eq:uh}, and DMD solution \eqref{JMH:eq:DMD_density}, and the bottom plot shows the DMD and FEM error \eqref{JMH:eq:DMD_density_error_MNF}.}
        \label{JMH:fig:mnf_snapshot_with_E}
    \end{minipage}
\end{figure}

We report results for the DMD model using several different values for the parameter $p$, which defines the number of the greatest singular values kept in the truncated SVD decomposition \eqref{JMH:eq:pseudo_inv_approx}.
Figure \ref{JMH:fig:mnf_flux_with_E} compares the boundary \JMHREV{photocurrent} from the manufactured solution and from the FEM and DMD models subject to the manufactured input \eqref{JMH:eq:MNF_input}. Figures \ref{JMH:fig:mnf_density_with_E} and \ref{JMH:fig:mnf_snapshot_with_E} show the carrier density from the manufactured solution and from the FEM and DMD models, with respect to time $t$ or position $x$, respectively. Figures \ref{JMH:fig:mnf_flux_no_E}, \ref{JMH:fig:mnf_density_no_E}, and \ref{JMH:fig:mnf_snapshot_no_E} show the same, but in the absence of an electric field.

\subsection{Verification test}\label{JMH:sec:num-ver}
\begin{figure}[htbp] 
   \centering
   \textbf{Training input and test input functions}\par\medskip
   \adjincludegraphics[width=0.48\textwidth, Clip={.0\width} {.0\height} {.0\width} {.1\height}]{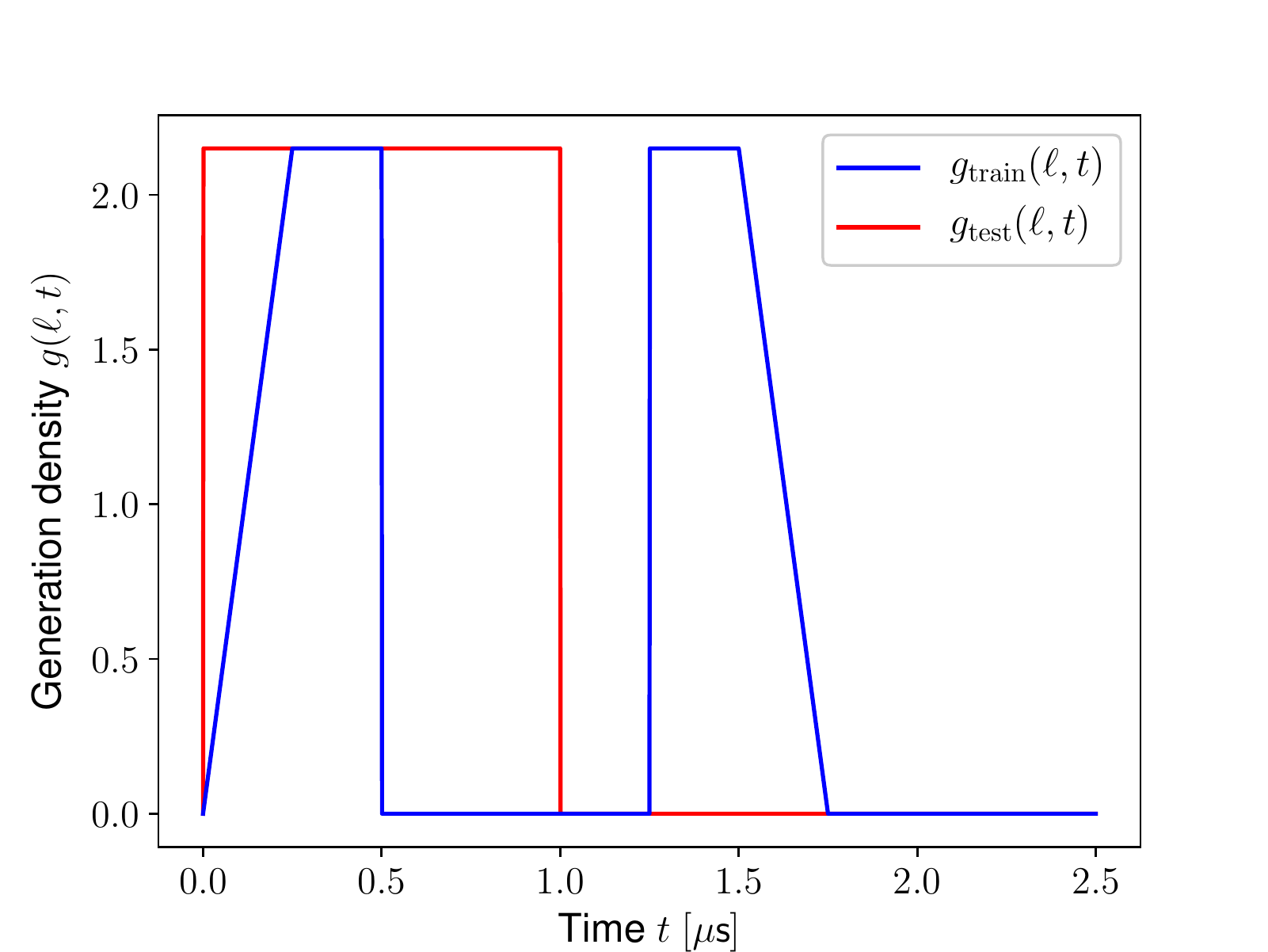}
   \caption{\scriptsize Generation density function $g_{\mathrm{train}}$ used to obtain the training set, and $g_{\mathrm{test}}$ used to verify the model performance. The square edges in the training input are included to excite a wider range of dynamic modes, which is due to the high-bandwidth content in the sharp transitions.}
   \label{JMH:fig:inputs}
\end{figure}
In this test we compare our \JMHPBB{compact} DMD model with the Axness-Kerr \cite{JMH:Axness_04_JAP} compact analytic model. We consider the case of a lightly doped diode \JMHREV{as described} in \cite[Section B, p.2650]{JMH:Axness_04_JAP} for which the length of the $N$-region equals 2 diffusion lengths (case $\xi_p = 2$ in \cite[Figure 3, p.2651]{JMH:Axness_04_JAP}.) 
\JMHPBB{The device is irradiated by a $\SI{1.0}{\micro\second}$ step pulse}
\begin{equation}\label{JMH:eq:test_input}
\JMHREV{g_{\mathrm{test}}}(x,t) =
\begin{cases}
\JMHREV{\widehat{g}} & \mbox{if} \ 0 \leq t \leq \SI{1.0}{\micro\second} \\[1ex]
0       & \mbox{otherwise}
\end{cases}
\end{equation}
\JMHPBB{with the value of $\widehat{g}$ rescaled as in \eqref{JMH:eq:scaledADE}.} This example from \cite{JMH:Axness_04_JAP} corresponds to the parameters in \ref{JMH:eq:scaledADE} with the exception of a few corrections to account for typographical errors in that paper. 

\JMHPBB{In contrast to the manufactured solution test where the desired input $g^{\mathrm{MNF}}(x,t)$ to the model is spatially irregular, now the target generation density, defined in \eqref{JMH:eq:test_input}, is spatially constant. As a result, the training input does not have to excite different regions of the device and can be chosen to be spatially constant as well. Thus, for this example} we choose the \textit{training input} $\JMHREV{g_{\mathrm{train}}}(x,t)$ to be a constant in space and a discontinuous piecewise linear in time function such that
\begin{equation}\label{JMH:eq:training_input}
\JMHREV{g_{\mathrm{train}}}(x,t) = 
\begin{cases}
0          & \mbox{if} \   4t    < \SI{0}{\micro\second}  \ \mbox{or} \  \   2 \leq 4t    < \SI{5}{\micro\second}  \ \mbox{or} \   4t \geq \SI{7}{\micro\second} \\[1ex]
\JMHPBB{\widehat{g}}(4t)   & \mbox{if} \ 0 \leq 4t    < \SI{1}{\micro\second} \\[1ex]
\JMHPBB{\widehat{g}}       & \mbox{if} \ 1 \leq 4t    < \SI{2}{\micro\second}   \ \mbox{or} \  5 \leq 4t    < \SI{6}{\micro\second} \\[1ex]
\JMHPBB{\widehat{g}}(7-4t) & \mbox{if} \ 6 \leq 4t    < \SI{7}{\micro\second} \\[1ex]
\end{cases},
\end{equation}
See Figure \ref{JMH:fig:inputs} for an illustration of $g_{\mathrm{train}}$ and $g_{\mathrm{test}}$. 

We then solve \eqref{JMH:eq:ADE} with the \JMHREV{source \eqref{JMH:eq:training_input}} and the parameters \eqref{JMH:eq:scaledADE} on a mesh $\mathcal{X}^h$ comprising 1024 uniform elements, and sample the solution in time at $\Delta t = \SI{0.0025}{\micro\second}$ increments. The compact DMD photocurrent model for the device is now defined according to \eqref{JMH:eq:DMD_model}.

\begin{remark}
Since the input is applied uniformly across the entire device (and thus the entire state space), we could reduce the input dimension of the DMD model \eqref{JMH:eq:DMD_model} from $N$ to $1$, however the higher dimensional input permits spatially irregular excitations such as localized radiation pulses or non-transversal plane waves \JMH{as} \JMHPBB{required for the manufactured solution test in Section \ref{JMH:sec:num-mnf}.}
\end{remark}

\begin{figure}
    \centering
    \textbf{Singular value decay}\par\medskip
	\adjincludegraphics[width=0.48\textwidth, Clip={.0\width} {.0\height} {.0\width} {.1\height}]{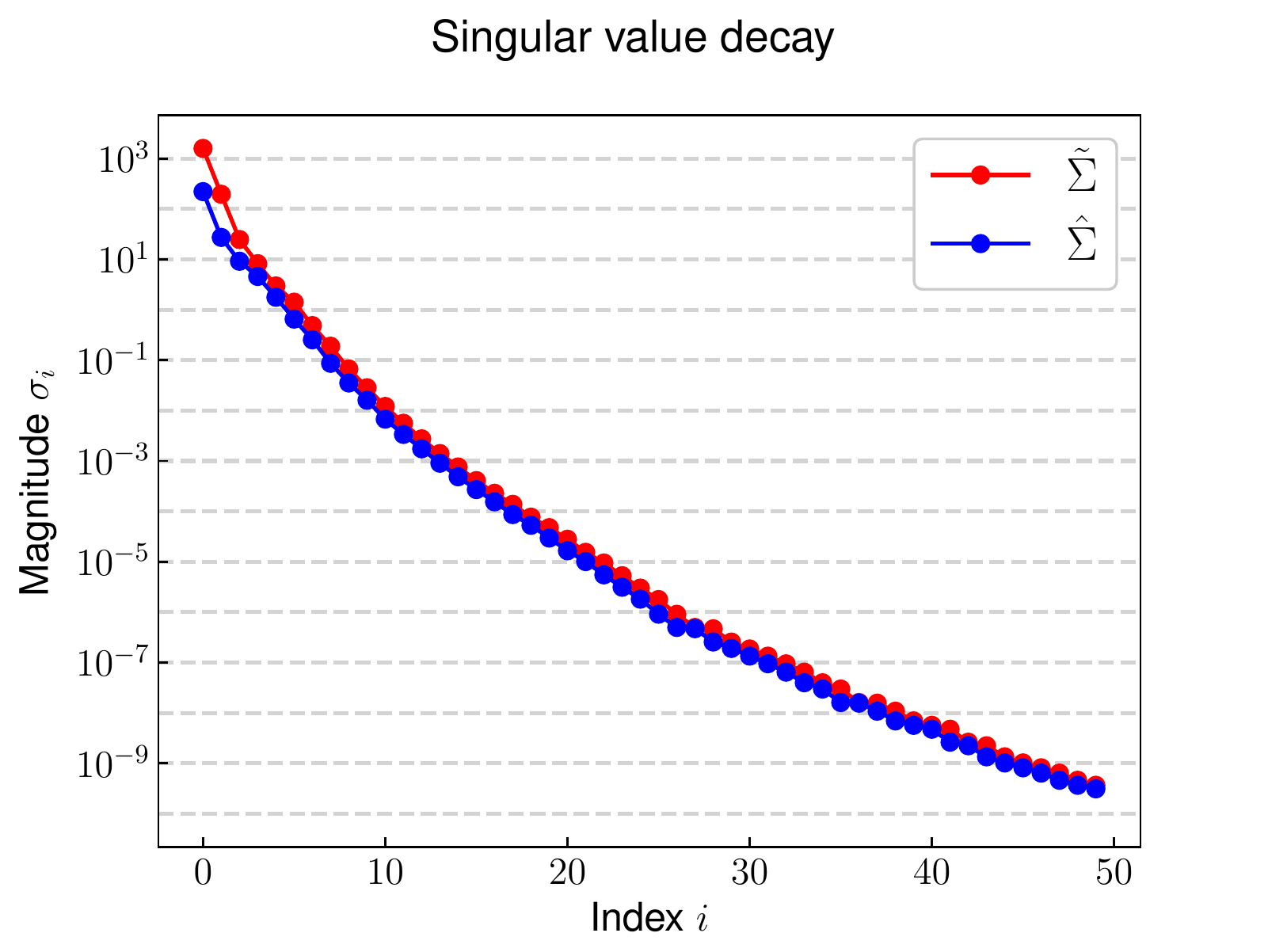}
    \caption{\scriptsize When listed from greatest to least, the singular values in the decomposition of the state-input sample matrix $S$ ($\tilde{\Sigma}$) and state sample matrix $X'$ ($\hat{\Sigma}$) demonstrate a nearly exponential decay in magnitude.}
    \label{JMH:fig:SVD_rolloff}
\end{figure}

To determine an appropriate dimension for the reduced-order DMD model, it is informative to inspect the relative magnitudes of the singular values from the decompositions of the sample matrices.
For the training input \eqref{JMH:eq:training_input}, Figure \ref{JMH:fig:SVD_rolloff} illustrates the magnitude roll-off in the singular values for the state and state-input sample matrix decompositions.
Observe that even for tight error thresholds, only a few modes are necessary to construct an accurate approximation of the state transition and input matrices.

We wish to emphasize that including too many modes in the reduced-order model often leads to a realization which is unstable over a long time horizon.
This phenomenon occurs due to unstable modes corresponding to small singular values. 
Including these low-magnitude modes will typically lead to a better fit for the state dynamics resulting from the training input over the original training time horizon, but may lead to model divergence over a longer time horizon, signifying an example of \textit{overfitting}.

Based on the \JMHREV{photocurrent} \JMH{\eqref{JMH:eq:FEM_flux_left}-\eqref{JMH:eq:FEM_flux_right}} and carrier density \eqref{JMH:eq:uh} derived from the finite-elements model and the approximate \JMHREV{photocurrent} \eqref{JMH:eq:DMD_flux} and approximate carrier density \eqref{JMH:eq:DMD_density} derived from the DMD model, we redefine the following error quantities:
\begin{align}
    E_J(t) &:= | J^\mathrm{DMD}_h(t) - J_h^\mathrm{FEM}(t) | \label{JMH:eq:DMD_flux_error} \\
    E_u(t) &:= | u^\mathrm{DMD}_h(t) - u_h^\mathrm{FEM}(t) | \label{JMH:eq:DMD_density_error}
\end{align}
which will facilitate the performance evaluation of the DMD model for typical, spatially uniform input functions. As before, the DMD model must be trained separately for each electric field strength.

\begin{figure}
    \centering
    \begin{minipage}{0.48\textwidth}
        \centering
        \textbf{FEM vs\JMH{.} DMD \JMH{f}lux \\ (training input, with electric field)}\par\medskip
        \includegraphics[width=\textwidth]{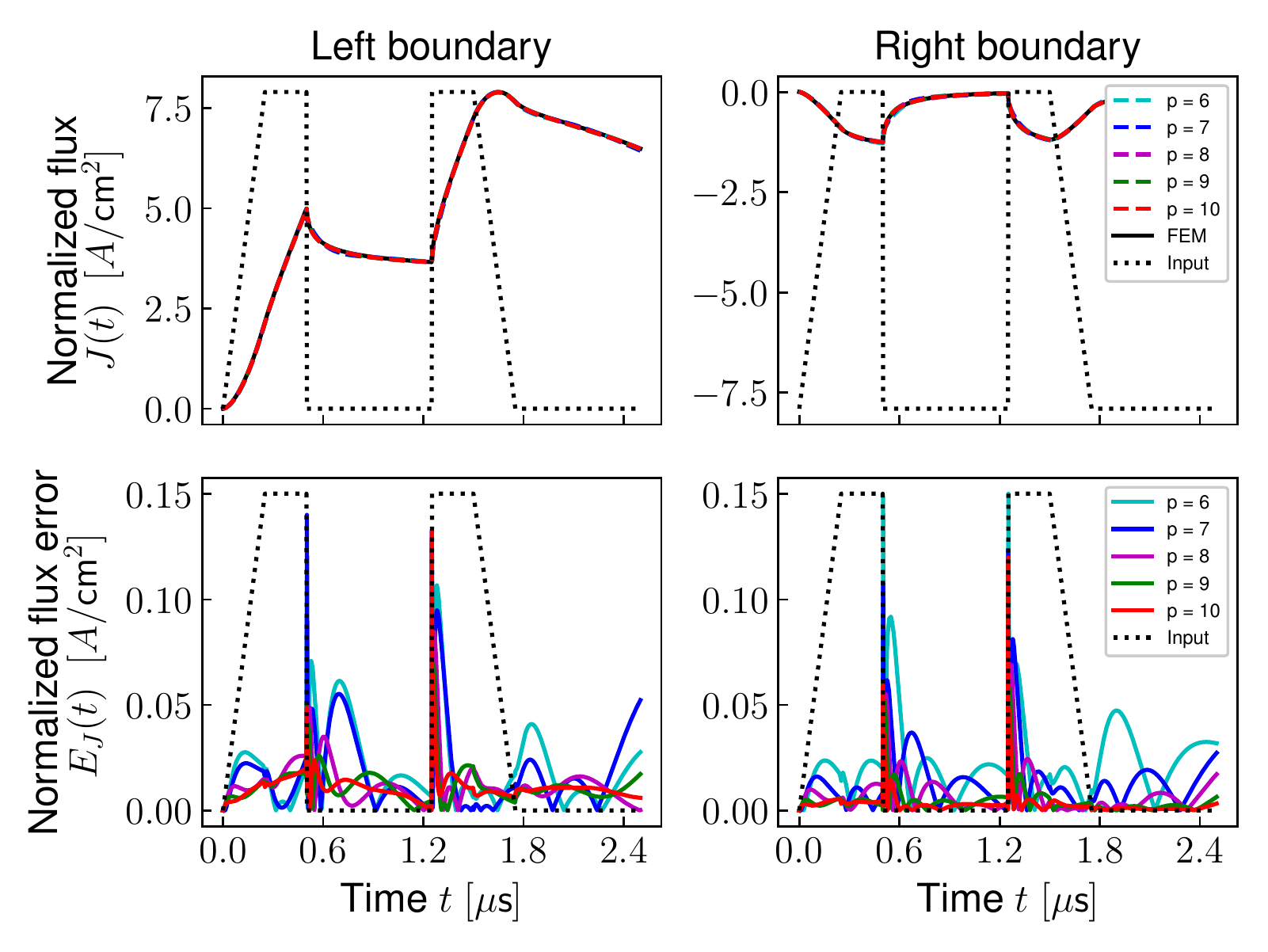}
        \caption{\scriptsize Normalized \JMHREV{photocurrent} due to the training input \eqref{JMH:eq:training_input} with an electric field \eqref{JMH:eq:scaledADE} applied. The top two plots show the FEM solution \JMH{\eqref{JMH:eq:FEM_flux_left}-\eqref{JMH:eq:FEM_flux_right}} and DMD solution \eqref{JMH:eq:DMD_flux}, and the bottom two plots show the DMD training error \eqref{JMH:eq:DMD_flux_error}. The left two plots correspond to the flux out of the left side of the \JMHREV{$N$-region}, and similarly on the right.}
        \label{JMH:fig:flux_train_with_E}
    \end{minipage}\hfill
    \begin{minipage}{0.48\textwidth}
        \centering
        \textbf{FEM vs\JMH{.} DMD \JMH{f}lux \\ (test input, with electric field)}\par\medskip
        \includegraphics[width=\textwidth]{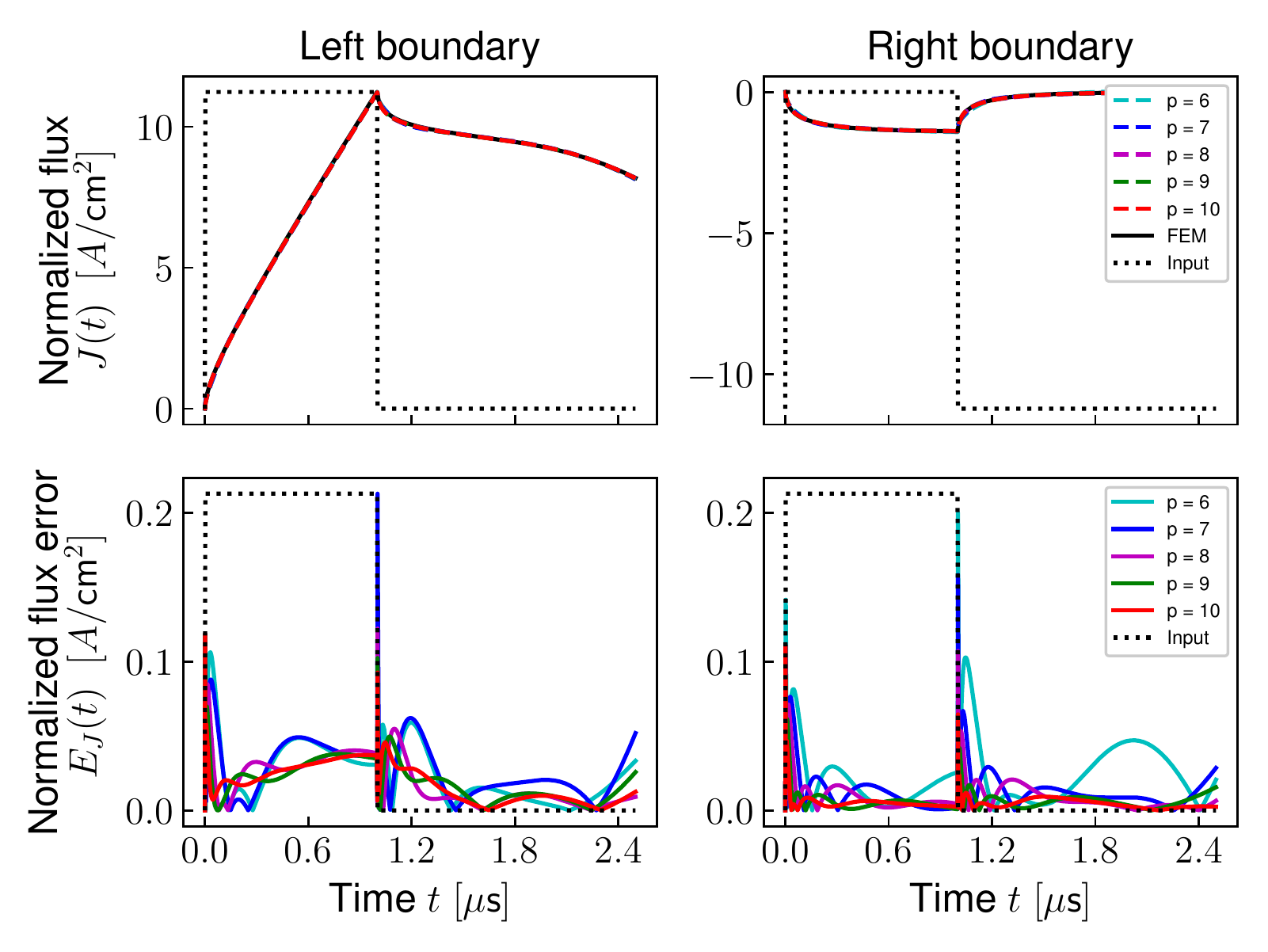}
        \caption{\scriptsize Normalized  \JMHREV{photocurrent} due to the test input \eqref{JMH:eq:test_input} with an electric field \eqref{JMH:eq:scaledADE} applied. The top two plots show the FEM solution \JMH{\eqref{JMH:eq:FEM_flux_left}-\eqref{JMH:eq:FEM_flux_right}} and DMD solution \eqref{JMH:eq:DMD_flux}, and the bottom two plots show the DMD training error \eqref{JMH:eq:DMD_flux_error}. The left two plots correspond to the flux out of the left side of the \JMHREV{$N$-region}, and similarly on the right.}
        \label{JMH:fig:flux_test_with_E}
    \end{minipage}
\end{figure}

\begin{figure}
    \centering
    \begin{minipage}{0.48\textwidth}
        \centering
        \textbf{FEM vs\JMH{.} DMD \JMH{f}lux \\ (training input, no electric field)}\par\medskip
        \includegraphics[width=\textwidth]{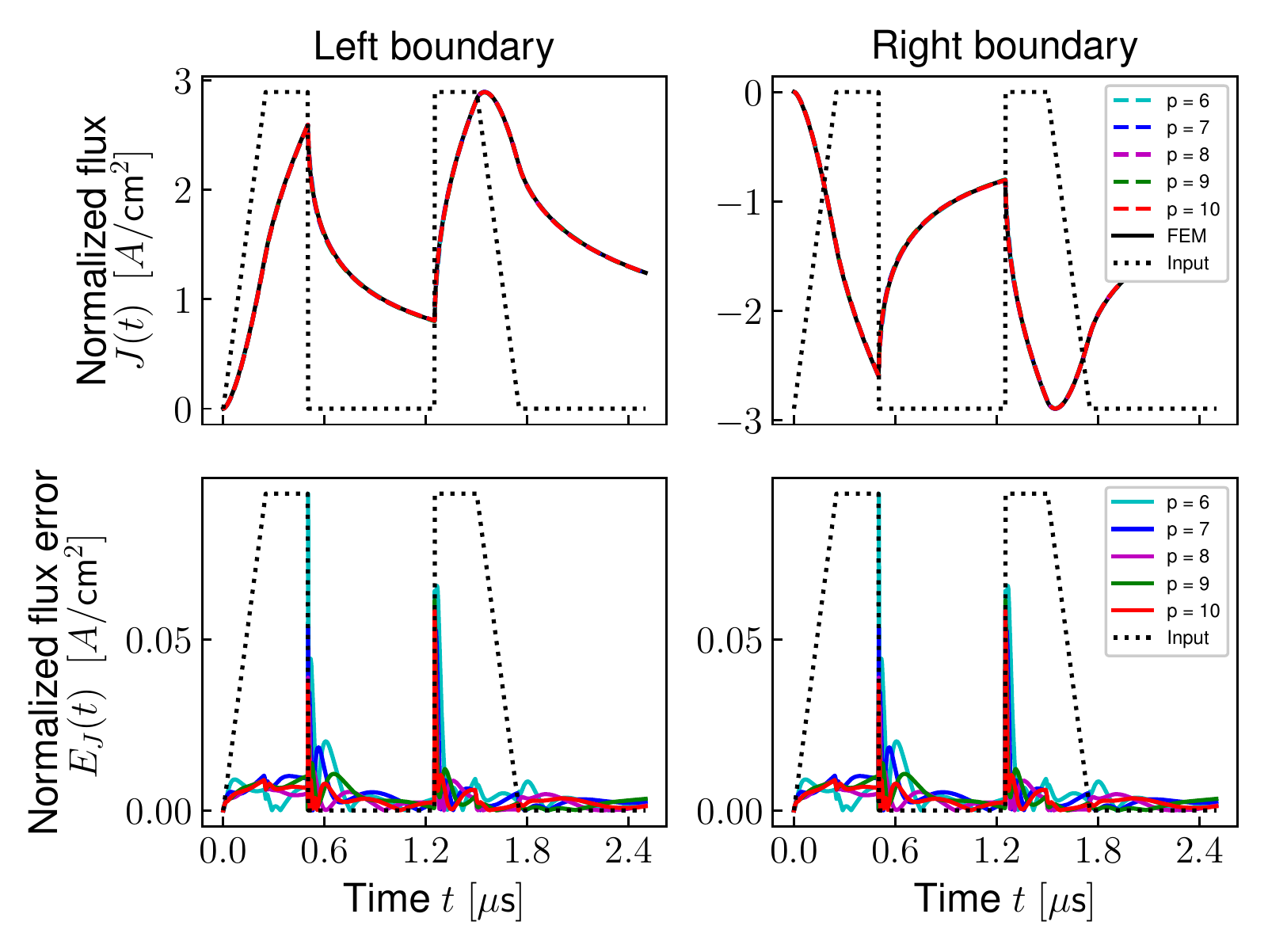}
        \caption{\scriptsize Normalized \JMHREV{photocurrent} due to the training input \eqref{JMH:eq:training_input} with no electric field applied. The top two plots show the FEM solution \JMH{\eqref{JMH:eq:FEM_flux_left}-\eqref{JMH:eq:FEM_flux_right}} and DMD solution \eqref{JMH:eq:DMD_flux}, and the bottom two plots show the DMD training error \eqref{JMH:eq:DMD_flux_error}. The left two plots correspond to the flux out of the left side of the \JMHREV{$N$-region}, and similarly on the right.}
        \label{JMH:fig:flux_train_no_E}
    \end{minipage}\hfill
    \begin{minipage}{0.48\textwidth}
        \centering
        \textbf{FEM vs\JMH{.} DMD \JMH{f}lux \\ (test input, no electric field)}\par\medskip
        \includegraphics[width=\textwidth]{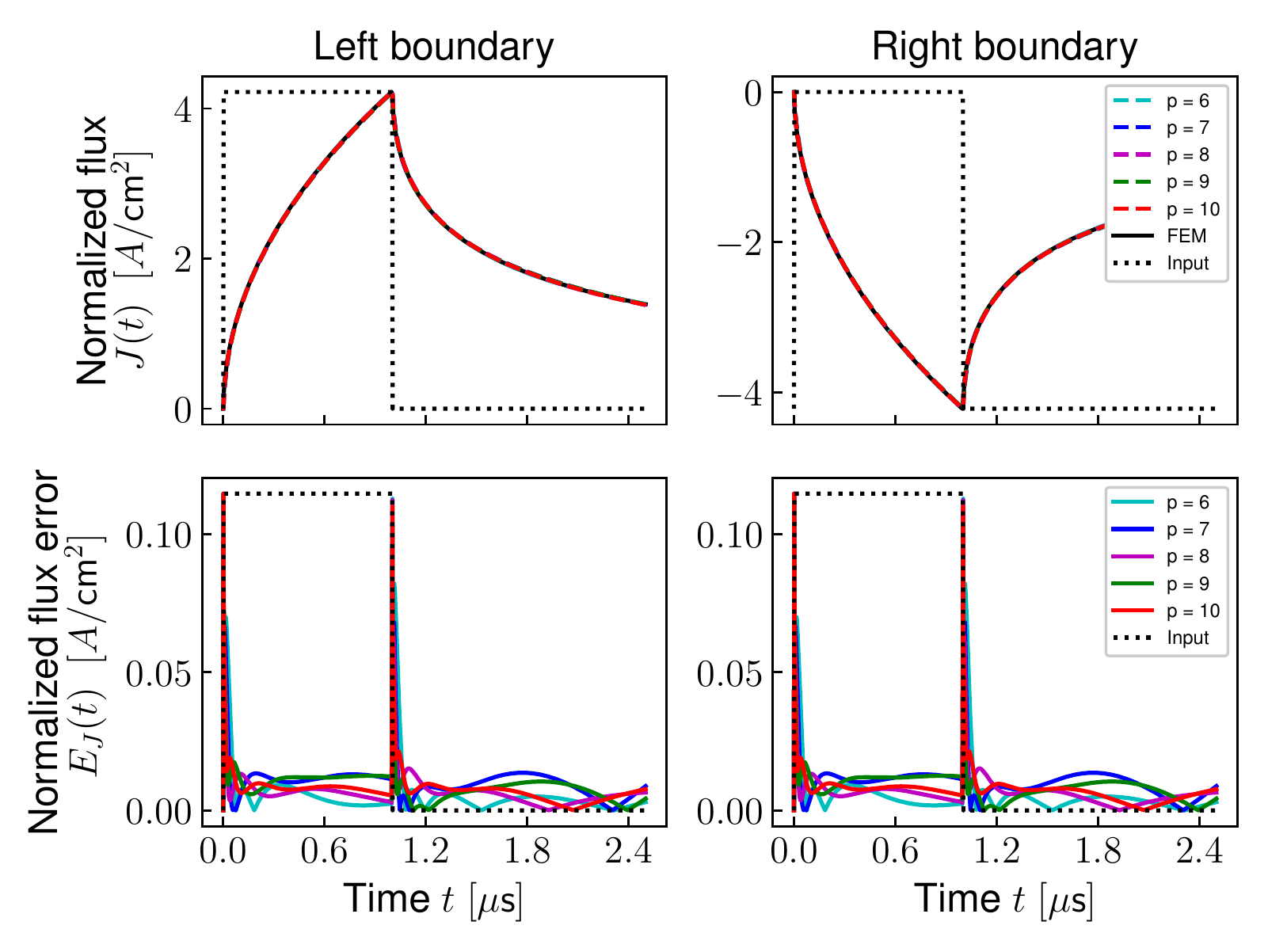}
        \caption{\scriptsize Normalized \JMHREV{photocurrent} due to the test input \eqref{JMH:eq:test_input} with no electric field applied. The top two plots show the FEM solution \JMH{\eqref{JMH:eq:FEM_flux_left}-\eqref{JMH:eq:FEM_flux_right}} and DMD solution \eqref{JMH:eq:DMD_flux}, and the bottom two plots show the DMD training error \eqref{JMH:eq:DMD_flux_error}. The left two plots correspond to the \JMHREV{photocurrent} out of the left side of the \JMHREV{$N$-region}, and similarly on the right.}
        \label{JMH:fig:flux_test_no_E}
    \end{minipage}
\end{figure}

\begin{figure}
    \centering
    \begin{minipage}{0.48\textwidth}
        \centering
        \textbf{FEM vs\JMH{.} DMD \JMH{d}ensity \\ (training input, with electric field)}\par\medskip
        \includegraphics[width=\textwidth]{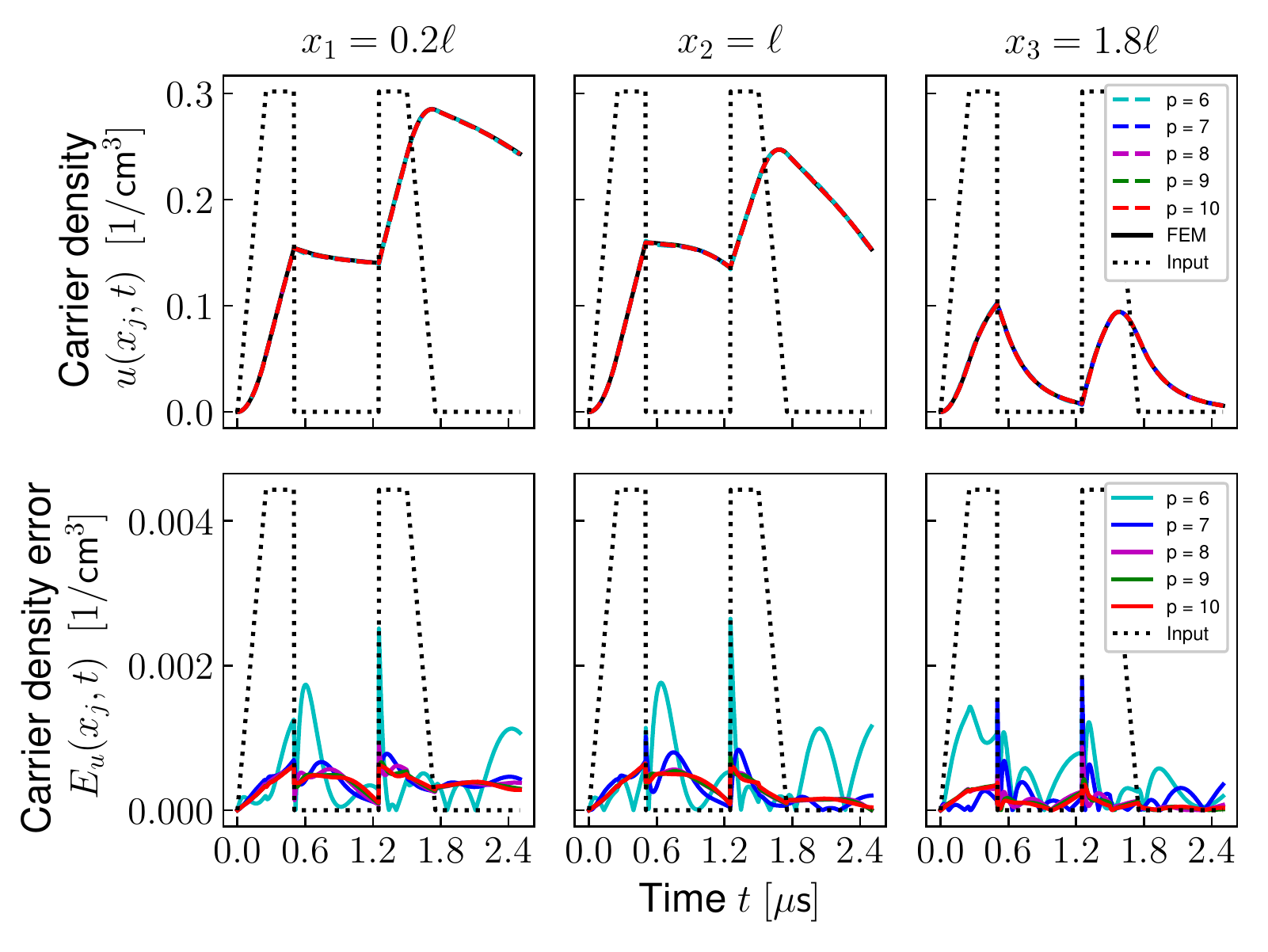}
        \caption{\scriptsize Simulated excess carrier density due to the training input \eqref{JMH:eq:training_input} with an electric field \eqref{JMH:eq:scaledADE} applied. The top three plots show the FEM solution \eqref{JMH:eq:uh} and DMD solution \eqref{JMH:eq:DMD_density}, and the bottom three plots show the DMD training error \eqref{JMH:eq:DMD_density_error}.}
        \label{JMH:fig:density_train_with_E}
    \end{minipage}\hfill
    \begin{minipage}{0.48\textwidth}
        \centering
        \textbf{FEM vs\JMH{.} DMD \JMH{d}ensity \\ (test input, with electric field)}\par\medskip
        \includegraphics[width=\textwidth]{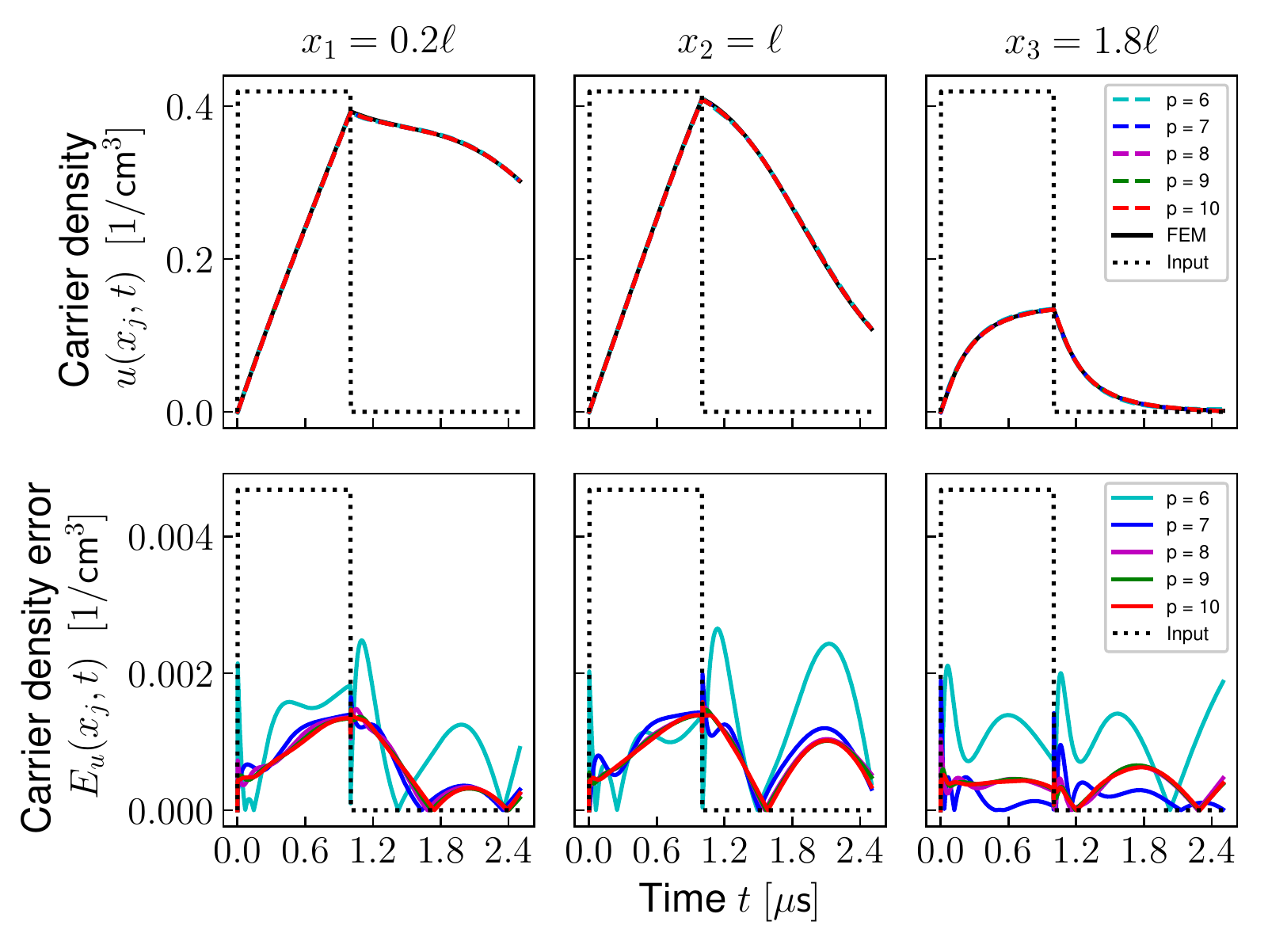}
        \caption{\scriptsize Simulated excess carrier density due to the test input \eqref{JMH:eq:test_input} with an electric field \eqref{JMH:eq:scaledADE} applied. The top three plots show the FEM solution \eqref{JMH:eq:uh} and DMD solution \eqref{JMH:eq:DMD_density}, and the bottom three plots show the DMD test error \eqref{JMH:eq:DMD_density_error}.}
        \label{JMH:fig:density_test_with_E}
    \end{minipage}
\end{figure}

Again, we show the results for the DMD model using multiple values for the parameter $p$. However, observe that for the case where the generation density is spatially uniform, fewer modes are necessary to achieve good performance than when the input magnitude varies along the length of the device, as in the manufactured solution test.
Figures \ref{JMH:fig:flux_train_with_E} and \ref{JMH:fig:flux_test_with_E} compare the boundary \JMHREV{photocurrent} produced by the FEM and DMD models for the training input \eqref{JMH:eq:training_input} and test input \eqref{JMH:eq:test_input}, respectively. Figures \ref{JMH:fig:flux_train_no_E} and \ref{JMH:fig:flux_test_no_E} show the same, but in the absence of an electric field. Figures \ref{JMH:fig:density_train_with_E} and \ref{JMH:fig:density_test_with_E} illustrate the simulated excess carrier density (i.e., the internal state) from the FEM and DMD models for both the training and test inputs. 

Several conclusions can be drawn from these results. First, the photocurrent plots at the left boundary of the $N$-region shown in Figures \ref{JMH:fig:flux_test_with_E} and \ref{JMH:fig:flux_test_no_E} are in an excellent agreement with the results reported in \cite[Figure 3, p.2651]{JMH:Axness_04_JAP} for $\xi_p=2$.
Second, the error plots on the bottom rows of the figures quantify the differences between the reference FEM solution and its DMD approximation as a function of the number $p$ of selected dynamical modes. These results reveal that, as expected, the error decreases with increase of the number of dynamic modes; however, even with just 6 modes selected the DMD photocurrent model yields excellent accuracy.
Overall, these results suggest that a data-driven approach is indeed a viable and effective alternative to traditional analytic model development that can be used to quickly develop accurate and computationally efficient photocurrent models directly from data.

\begin{figure}
    \centering
    \begin{minipage}{0.48\textwidth}
        \centering
        \textbf{FEM vs\JMH{.} DMD \JMH{d}ensity \\ (training input, no electric field)}\par\medskip
        \includegraphics[width=\textwidth]{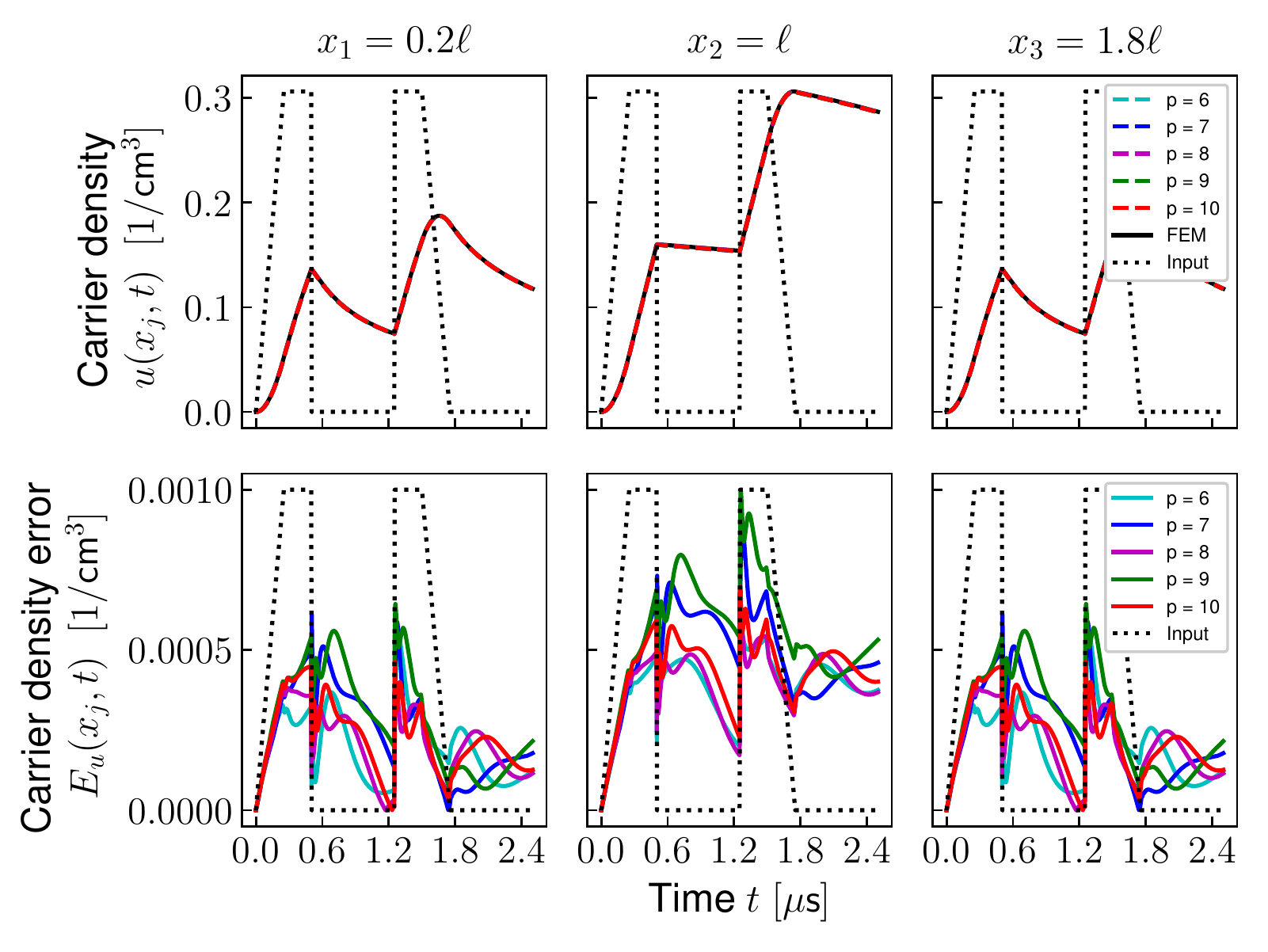}
        \caption{\scriptsize Simulated excess carrier density due to the training input \eqref{JMH:eq:training_input} with no electric field applied. The top three plots show the FEM solution \eqref{JMH:eq:uh} and DMD solution \eqref{JMH:eq:DMD_density}, and the bottom three plots show the DMD training error \eqref{JMH:eq:DMD_density_error}.}
        \label{JMH:fig:density_train_no_E}
    \end{minipage}\hfill
    \begin{minipage}{0.48\textwidth}
        \centering
        \textbf{FEM vs\JMH{.} DMD \JMH{d}ensity \\ (test input, no electric field)}\par\medskip
        \includegraphics[width=\textwidth]{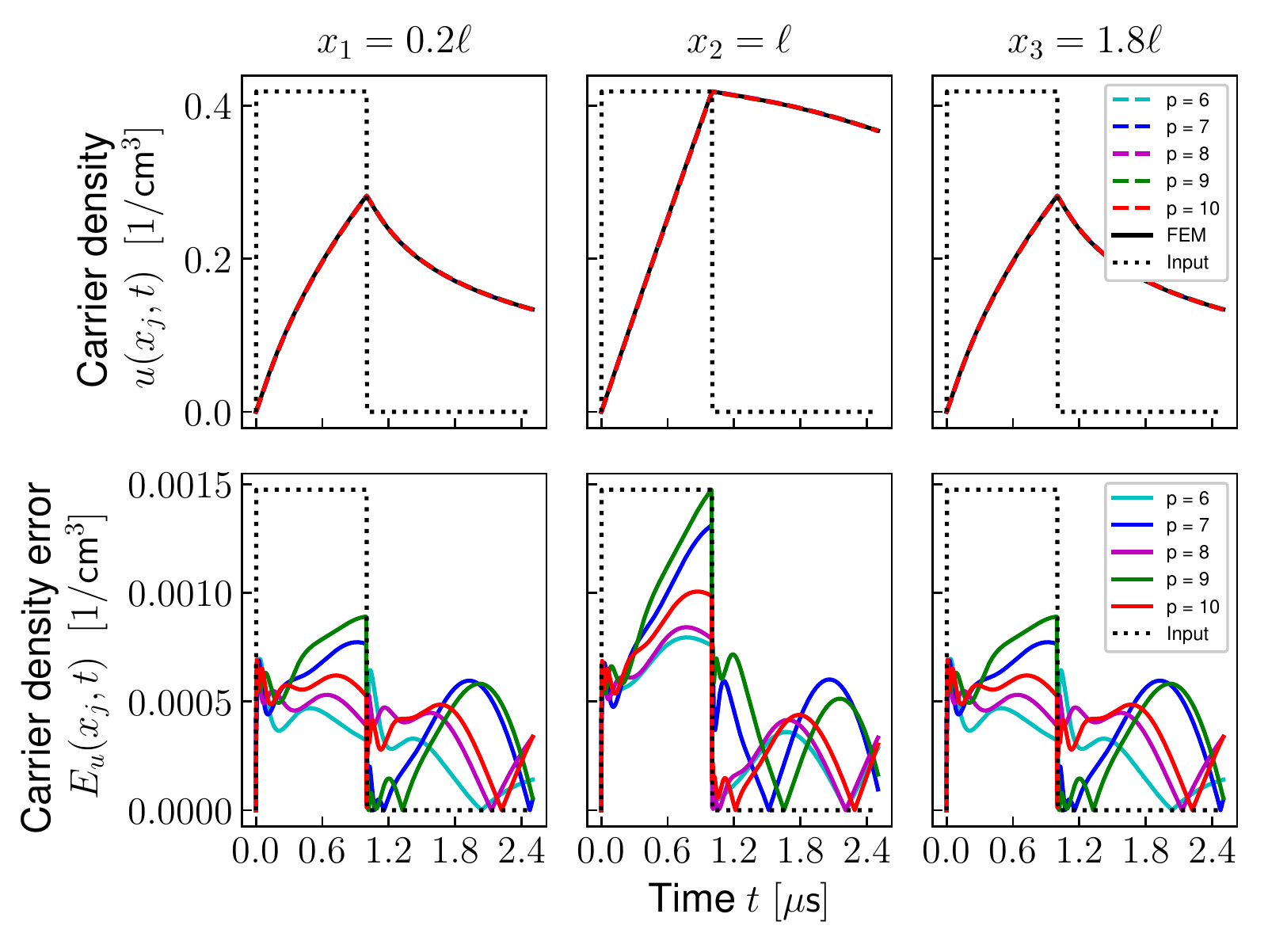}
        \caption{\scriptsize Simulated excess carrier density due to the test input \eqref{JMH:eq:test_input} with no electric field applied. The top three plots show the FEM solution \eqref{JMH:eq:uh} and DMD solution \eqref{JMH:eq:DMD_density}, and the bottom three plots show the DMD test error \eqref{JMH:eq:DMD_density_error}.}
        \label{JMH:fig:density_test_no_E}
    \end{minipage}
\end{figure}


\section{Conclusions}\label{JMH:sec:concl}
We have developed a compact data-driven \JMHPBB{delayed} photocurrent model \JMHPBB{given by a} \JMH{low-dimensional} \JMHPBB{discrete-time dynamical system, which approximates the flow map of the Ambipolar Diffusion Equation. 
To obtain the approximate flow map we use the Ambipolar Diffusion Equation to reconstruct numerically the internal state of the device, which is not directly observable through a laboratory measurement, and then apply Dynamic Mode Decomposition to the simulated internal state \JMH{samples}.
In \JMH{doing so} physics knowledge is incorporated into the model development, which allows us to obtain models from \emph{sparse data sets} that accurately approximate the dynamics of the excess carrier density.
This in turn allows us to accurately estimate the induced current at the device boundaries, which is the quantity required for circuit simulations.}

Our results confirm that such physics-aware data-driven models are a viable alternative to traditional analytic compact models that use simplified analytic solutions of the governing equations and often must undergo recalibration and/or redevelopment to include new physics effects. 

Our future work will consider extension of the approach \JMHPBB{to include an additional parameter identification step, and to} the fully coupled DDE system \eqref{JMH:eq:DD1}-\eqref{JMH:eq:DD3}. \JMHPBB{The latter will allow us to model the total photocurrent in the device and eliminate the need to split it into three separate regions.}

Applying nonlinear observable functions to the state of the DDE would allow us to model the nonlinear problem using the same DMD algorithm, requiring little to no additional computational cost for running and training the DMD model (besides applying the nonlinearities to the measurement data, which is inexpensive). We also plan to incorporate and test our models in circuit simulators to demonstrate their utility for circuit design and analysis tasks.

\section*{Acknowledgments}
This work was supported by the Sandia National Laboratories (SNL) Laboratory-directed Research and Development (LDRD) program, and the U.S. Department of Energy, Office of Science, Office of Advanced Scientific Computing Research under Award Number DE-SC-0000230927 and under the Collaboratory on Mathematics and Physics-Informed Learning Machines for Multiscale and Multiphysics Problems (PhILMs) project. 
The work of J. Hanson was performed as part of Sandia Summer Student Program Internship at the Computer Science Research Institute (CSRI).

Sandia National Laboratories is a multimission laboratory managed and operated by National Technology and Engineering Solutions of Sandia, LLC., a wholly owned subsidiary of Honeywell International, Inc., for the U.S. Department of Energys National Nuclear Security Administration contract number DE-NA0003525. This paper describes objective technical results and analysis. Any subjective views or opinions that might be expressed in the paper do not necessarily represent the views of the U.S. Department of Energy or the United States Government.

This work benefited from numerous discussions and interactions with our colleagues Eric Keiter, Suzey Gao and Larry Musson who shared their expertise with compact model development and TCAD simulations. We are grateful for their help and advice throughout the preparation of this paper.

\bibliographystyle{siam}
\bibliography{JoshuaHanson}


\end{document}